\documentclass[journal,onecolumn]{IEEEtran}
\IEEEoverridecommandlockouts
\usepackage{cite}
\usepackage{amsmath,amssymb,amsfonts}
\usepackage{algorithmicx}
\usepackage{graphicx}
\usepackage{textcomp}

\usepackage{color}
\usepackage{algorithm}
\usepackage{algpseudocode}
\usepackage{subcaption}

\usepackage{amsthm}

\newtheorem{theorem}{Theorem}

\newtheorem{lemma}[theorem]{Lemma}
\newtheorem{proposition}[theorem]{Proposition}
\newtheorem{aug}{Treeplication Augmentation$\!$}
\newenvironment{augment}{\begin{aug}\hspace*{-1ex}{\bf}}{\end{aug}}
\newtheorem*{augn*}{Augmentation by replication}
\newenvironment{augmentn}{\begin{augn*}\hspace*{-1ex}{\bf}}{\end{augn*}}

\newtheorem{defn}[theorem]{Definition}

\newcommand{\kfrag}{k}
\newcommand{\fragsz}{D}
\newcommand{\nvert}{n}
\newcommand{\tre}{T}
\newcommand{\Qm}{Q^{\star}}
\newcommand{\mwgt}{n}
\newcommand{\nodeset}{\mathcal{N}}
\newcommand{\nodesset}{\mathcal{N}_{A}}
\newcommand{\vchosen}{\sigma}
\newcommand{\vparent}{\rho}
\newcommand{\vsibling}{\sigma'}
\newcommand{\pP}{\pi}

\definecolor{auburn}{rgb}{0.43, 0.21, 0.1}
\newcommand\itw[1]{{\color{black}#1}}

\newcommand\yc[1]{{\color{black}#1}}

\def\BibTeX{{\rm B\kern-.05em{\sc i\kern-.025em b}\kern-.08em
    T\kern-.1667em\lower.7ex\hbox{E}\kern-.125emX}}
\begin{document}

\title{Treeplication: An Erasure Code for Distributed Full Recovery under the Random Multiset Channel}

\author{\large
	Michael~Gandelman and
	Yuval~Cassuto,~\IEEEmembership{Senior Member,~IEEE}
\vspace{-4ex}
\thanks{Michael Gandelman and Yuval Cassuto are with the Viterbi Department of Electrical Engineering, Technion -- Israel Institute of Technology, Haifa Israel (emails: michaelgandelman@gmail.com, ycassuto@ee.technion.ac.il).}
\thanks{This work was supported in part by the Israel Science Foundation, and in part by the US-Israel Binational Science Foundation.}
\thanks{Part of the results in the paper were presented at the IEEE Information Theory Workshop, November 2018.}	
}

\maketitle

\begin{abstract}

This paper presents a new erasure code called Treeplication designed for distributed recovery of the full information word, while most prior work in coding for distributed storage only supports distributed repair of individual symbols. A Treeplication code for $k$ information symbols is defined on a binary tree with $2k-1$ vertices, along with a distribution for selecting code symbols from the tree layers. We analyze and optimize the code under a random-multiset model, which captures the system property that the nodes available for recovery are drawn randomly from the nodes storing the code symbols. Treeplication codes are shown to have full-recovery communication-cost comparable to replication, while offering much better recoverability.

\end{abstract}

\IEEEpeerreviewmaketitle

\section{Introduction}
To design a distributed storage system, one has to balance the costs of storage and communications to reach a certain degree of data availability. Traditional maximum distance separable (MDS) codes, such as Reed Solomon codes~\cite{RS:60} and array codes~\cite{BlaumM:98b}, minimize the storage cost with no regard to communication costs if symbols are distributed across nodes in a network. To economize communication, a new field of coding theory -- coding for distributed storage -- has emerged, contributing many new codes with many advantages in distributed settings. The principal problem addressed by codes for distributed storage is the {\em repair} of lost symbols with 1) little communication from nodes storing the other symbols (regenerating codes~\cite{Dimakis,RasShaKum_pm,YeBarg:17,Tian,determinant_regen}), or 2) communication from few other nodes (locally repairable codes (LRC)~\cite{Gopalan,pklk,oggier2011self,TamoBarg:14,pd,lrc_matroid,SilRawKoyVis}). Going beyond the repair problem, in this paper we consider distributed recovery of the full information word, where each information symbol is recovered by a different available node, with little communication among the available nodes. This scenario, which we call {\em distributed full recovery} for short, is useful in distributed systems that cannot tolerate the high complexity and latency of centralized decoding when accessing the information symbols in parallel. An important use of such systems is {\em map-reduce} distributed computing~\cite{MapReduce}, where large data units\footnote{The term data unit will henceforth replace the term information word.} are processed in parallel by multiple machines, each working on one fragment\footnote{The term fragment will henceforth replace the term symbol.} of the data unit. Both regenerating codes and LRC codes assume {\em centralized} full recovery, where the former expresses this as the ``data collector'' function, and in the latter the code minimum distance specifies the centralized full-recovery capabilities.          
   
When using {\em replication}, one gets distributed full recovery trivially, because every node storing a data fragment can access it locally without any communication. However, replication fails full recovery if even one data fragment is missing from the set of available nodes; hence for adequate full-recovery availability many replicas need to be deployed in many nodes, entailing steep equipment costs. Using an {\em erasure code} instead of replication will attain the same full-recovery availability with fewer deployed fragments (some of which are parity fragments), while requiring some nodes to communicate in order to recover a data fragment from the parity fragment they store. The objective of this paper is to develop such an erasure code, in which the full-recovery communication costs are comparable to replication, but with a much better full-recovery availability.        

Throughout the paper a data unit is divided to $\kfrag$ data fragments, and encoded by an erasure code to $m$ code fragments, each stored in a distinct node. Some of the code fragments are systematic data fragments and others are parity fragments. As discussed above, we define the decoding operation to be distributed full recovery, namely, $\kfrag$ nodes out of a set of $\nvert$ {\em available nodes} each recovers a different one of the $\kfrag$ data fragments. The set of $\nvert$ available nodes is drawn randomly from the set of $m$ nodes storing the data unit's code fragments, and we assume that the difference $m-\nvert$ can be large, that is, the available node set has a highly punctured version of the length-$m$ codeword. We defer the formal definition of the random drawing model to Section~\ref{subsec:rand_multiset}. Note that replication is a special case of this setup, in which distributed full recovery succeeds if and only if each of the $\kfrag$ data fragments is present (at least once) in the $\nvert$ available nodes. That said, replication has especially poor full-recovery probability in randomized settings, due to the {\em coupon collector problem}~\cite{Mitzenmacher} requiring to draw many fragments ($\Theta(\kfrag \log \kfrag)$) to succeed with high probability. One should think about the parameter $\kfrag$ as the degree of parallelization (number of nodes) needed to process a data unit.  

The erasure code we propose for communication-efficient distributed full recovery is called {\em Treeplication}, owing to its code fragments being generated from a binary-tree structure. A Treeplication code is defined over a perfect binary tree with $\kfrag$ leaf vertices and $2\kfrag-1$ vertices in total. Each leaf vertex represents a data fragment, and each non-leaf vertex represents the bitwise exclusive-or (XOR) of its two children. The tree structure allows on the one hand localized erasure correction that improves the communication efficiency of the code compared to existing erasure codes, and on the other hand spans large-degree parity symbols that improve decodability compared to replication. Once the code structure is set, the main contributions of this paper are showing how to select fragments from the tree to maximize the code performance under randomized node availability, and providing exact analytic evaluations of the code's decodability and communication-cost performance. After formal definition of the drawing model and of the Treeplication code, our results can be divided into three parts. The first part (Sections~\ref{sec:uniform},~\ref{sec:nonuniform}) focuses on the full-recovery decodability of Treeplication, where we provide analytic expressions for the decodability probability and an efficient algorithm to find the optimal fragment selection distribution given the size $\nvert$ of the available set. At the end of this part we show the advantage of Treeplication over replication in terms of full-recoverability, in particular, replication requires 60\% more available fragments than Treeplication in order to achieve the same probability of full-recovery decodability. In the second part (Section~\ref{RC}), we study the communication cost of Treeplication, defined as the total number of fragments communicated in a full recovery of a decodable fragment subset. The main results in this section are 1) an algorithm that finds the recovery schedule with minimal total communication cost, and 2) an analytic expression for the full distribution of the total communication cost under any fragment selection distribution. In Treeplication coding full recovery has the attractive properties that (like in replication) only $\kfrag$ nodes (out of the $\nvert$ available nodes) participate in recovery, and each participating node has to send its fragment to at most one other node.  Using the derived communication-cost distribution we show that the average communication cost of the optimal Treeplication codes from Section~\ref{sec:nonuniform} is lower than MDS codes by more than an order of magnitude. Finally, in the third part of the paper (Sections~\ref{sec:health},~\ref{sec:augment}) we move to study Treeplication in dynamic settings where the available fragments of data units change in time. To that end in Section~\ref{sec:health} we propose a measure we call {\em tree health} that provides a tractable way to evaluate and compare the robustness of Treeplication-coded data units to future events of fragment loss. Then in Section~\ref{sec:augment} we discuss methods to {\em augment} a Treeplication data unit with additional fragments, while having access to only a small subset of the nodes storing the data unit.

At a high level, the Treeplication scheme goes the opposite direction to most recent works on distributed-storage erasure coding: instead of taking an erasure code and making it more ``access friendly", we take the replication scheme and gracefully make it more ``storage-cost friendly". That said, it is possible that known regenerating and/or LRC codes (or improvements thereof) can be used toward efficient distributed full recovery. In addition to typically requiring large fragment sizes, the challenges in using current regenerating codes are that high repair efficiencies require more nodes to perform recovery, and that multiple simultaneous repairs (as needed for full recovery) require nodes to send their fragments to many other nodes, demanding high upload bandwidths. The main challenge of using LRC codes (and their relative availability codes~\cite{pamies}) is to obtain multiple simultaneous repair sets in heavily punctured recovery instances.  

The structure of Treeplication is inspired by the similar structure of the fountain code proposed in~\cite{Bundle}, but, among several key differences, our codes are designed for optimal performance in fixed values of $\kfrag$, while the results in the prior work are asymptotic.

\subsection{The random-multiset model}\label{subsec:rand_multiset}
In classical distributed-storage erasure coding, one generates a codeword of $m$ code fragments, out of which $n$ fragments are available for decoding while the other $m-n$ are unavailable (viewed as erasures). It has been universally assumed that the $m$ code fragments are {\em all distinct}, and thus the $n$ available fragments form {\em a subset} of the code fragments. In this paper we lift this assumption, and generate $m$ code fragments {\em with multiplicity} from $M$ distinct code fragments. This makes the $n$ available fragments {\em a multiset} of the $M$ distinct code fragments. By doing so, we can fix the $M$ distinct code fragments to have some desired structure, which in this paper contributes to efficient distributed full recovery of data units. An extreme case of this approach is standard replication, where $M=\kfrag$, and the distinct code fragments are simply the $\kfrag$ data fragments. The $M$ distinct code fragments of replication enjoy the very convenient structure that decoding is a trivial no-operation, but at the cost of many non-decodable multisets even for fairly large $n$ (the coupon collector's problem). In the Treeplication coding scheme we have $M=2\kfrag-1$ distinct code fragments in a tree structure that offers decoding efficiency, but with much better decodability performance than in replication. An important part of the Treeplication code design is the {\em fragment selection distribution}, specifying how the $m$ code fragments are drawn from the $M$ distinct code fragments. In addition, for probabilistic analysis of multiset decodability we need to define the {\em erasure channel} selecting a multiset of $n$ available fragments from the multiset of the $m$ generated ones. To make for a simpler and cleaner analysis, we merge the fragment selection distribution and the erasure channel into one {\em random-multiset model}, specifying how the available $n$-multiset is drawn from the $M$ distinct code fragments, without need to deal with an extra parameter $m$. In the paper we use two main random-multiset models: in the uniform model (used in Section~\ref{sec:uniform}) we draw with replacement $\nvert$ fragments from the $M=2\kfrag-1$ tree vertices of Treeplication; in the non-uniform model (used in Section~\ref{sec:nonuniform}) we draw with replacement $\nvert_i$ fragments from the vertices of layer $i$ of the Treeplication tree.     
\itw{
\section{Treeplication Coding}\label{sec:coding}
A distributed storage system stores {\em data units} across storage nodes. Each data unit is broken to $\kfrag$ {\em data fragments}, each of size $\fragsz$. Throughout the paper we assume that $\kfrag=2^s$, for some integer $s$. To encode the data unit in the storage system, we use a {\em binary tree} of depth $s$. The tree has $2^s$ leaf vertices representing the data fragments, and a total of $2^{s+1}-1$ vertices. Note that including the root there are $d\triangleq s+1$ {\em layers} in the tree. The layers are numbered from bottom to root, thus for $i\in\{1,\ldots,d\}$, layer $i$ has $2^{d-i}$ vertices. In the sequel we use $\tre_d$ to denote this tree, and $\tre_{\ell}$ refers to one of $\tre_d$'s subtrees with $2^{\ell}-1$ vertices in layers $\{1,\ldots,\ell\}$; both $\tre_d$ and its $\tre_{\ell}$ subtrees are complete binary trees. The interpretation of the tree is that each vertex represents a {\em code fragment}: starting from level $i=2$ each code fragment is the bit-wise exclusive-or (XOR) of its two children, and the leaves at level $i=1$ are the ``pure" data fragments also called {\em systematic code fragments}. An example of this tree representation is given in Fig.~\ref{fig:treeplication} for the case $\kfrag=8$.
\begin{figure}[htbp]
\centerline{\includegraphics[width=260pt]{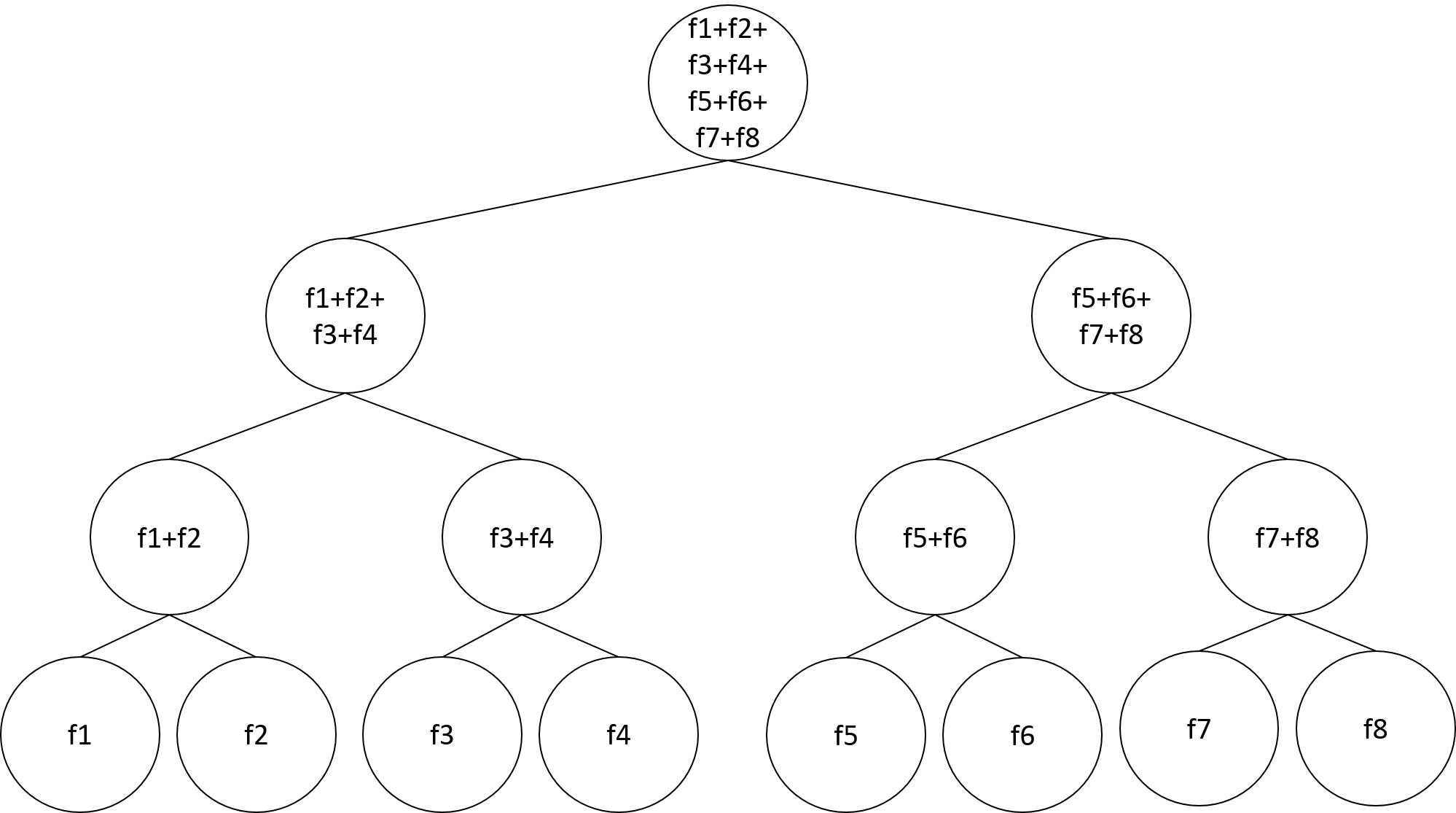}}
\caption{Tree representation of Treeplication coding for $\kfrag=8$ data fragments ($s=3$,$d=4$).}
\label{fig:treeplication}
\end{figure}

The following is a central definition for the analysis and design of Treeplication codes.
\begin{defn}
A \textbf{Treeplication subset} for a tree $\tre_{\ell}$ is defined as a subset of the vertices of $\tre_{\ell}$. 
\end{defn}
A Treeplication subset represents the code fragments available in system nodes for useful operations such as data-fragment recovery. In the sequel we refer to a vertex and its associated code fragment interchangeably. Moreover, when there is no risk of confusion, we refer to a Treeplication subset simply as {\em a subset}.
 \begin{defn}
A Treeplication subset of $\tre_{\ell}$ is said to be \textbf{decodable} if the corresponding code fragments are sufficient to recover the $\kfrag'\triangleq 2^{\ell-1}$ data fragments. 
\end{defn} 
Clearly, a decodable subset must have at least $\kfrag'$ vertices, and also all subsets of size greater than $2\kfrag'-3$ are decodable. Between $\kfrag'$ and $2\kfrag'-3$, decodability depends on the particular subset available for decoding. (Viewed as an erasure code with block length $2\kfrag'-1$, $\tre_{\ell}$ can correct any single erasure, and many combinations of between $2$ and $\kfrag'-1$ erasures, but not more than $\kfrag'-1$ erasures.) An example of a decodable subset of $\tre_{4}$ ($\kfrag=8$) with exactly $\kfrag$ vertices is illustrated in Fig.~\ref{fig:decsubset}. Similarly, Fig.~\ref{fig:decsubset2} illustrates a decodable subset with more than $\kfrag$ vertices. From linearity of the code, a decodable subset with more than $\kfrag$ vertices is redundant, and $\kfrag$ vertices from the subset are always sufficient to decode the data unit. For instance, in the example presented in Fig.~\ref{fig:decsubset2} there are $10>8$ vertices in the decodable subset, meaning that $2$ elements of the subset may be discarded without affecting decodability (e.g. those corresponding to $f3$ and $f7+f8$, or those corresponding to $f1+f2+f3+f4$ and the root of the tree). The following lemmas further characterize decodable subsets. 
\begin{lemma}\label{lem:subtree_decodable}
If a subset of $\tre_{\ell}$ is decodable, then at least one immediate subtree: $\tre_{\ell-1}$ (left) or $\tre'_{\ell-1}$ (right) is decodable with only vertices from its subtree.
\end{lemma}
\begin{IEEEproof}
Since the subtrees $\tre_{\ell-1}$,$\tre'_{\ell-1}$ are disjoint in their XOR arguments, having both non-decodable internally would mean that two additional code fragments are needed. This is a contradiction because there is only the root as a potential extra code fragment.
\end{IEEEproof}

\begin{figure}[htbp]
\centerline{\includegraphics[width=200pt]{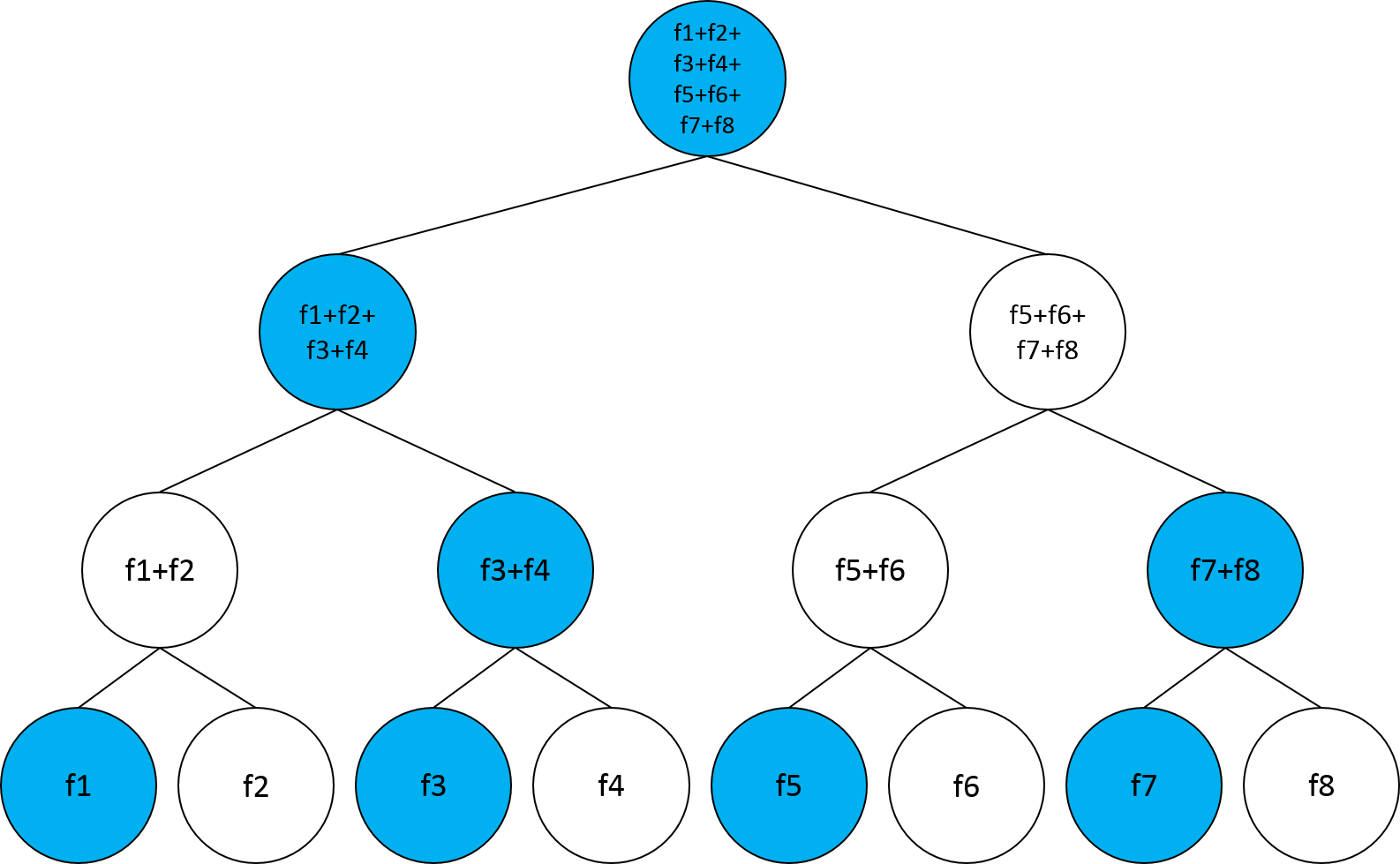}}
\caption{Example of a decodable Treeplication subset with exactly $\kfrag=8$ vertices (the filled vertices).}
\label{fig:decsubset}
\end{figure}

\begin{figure}[htbp]
\centerline{\includegraphics[width=200pt]{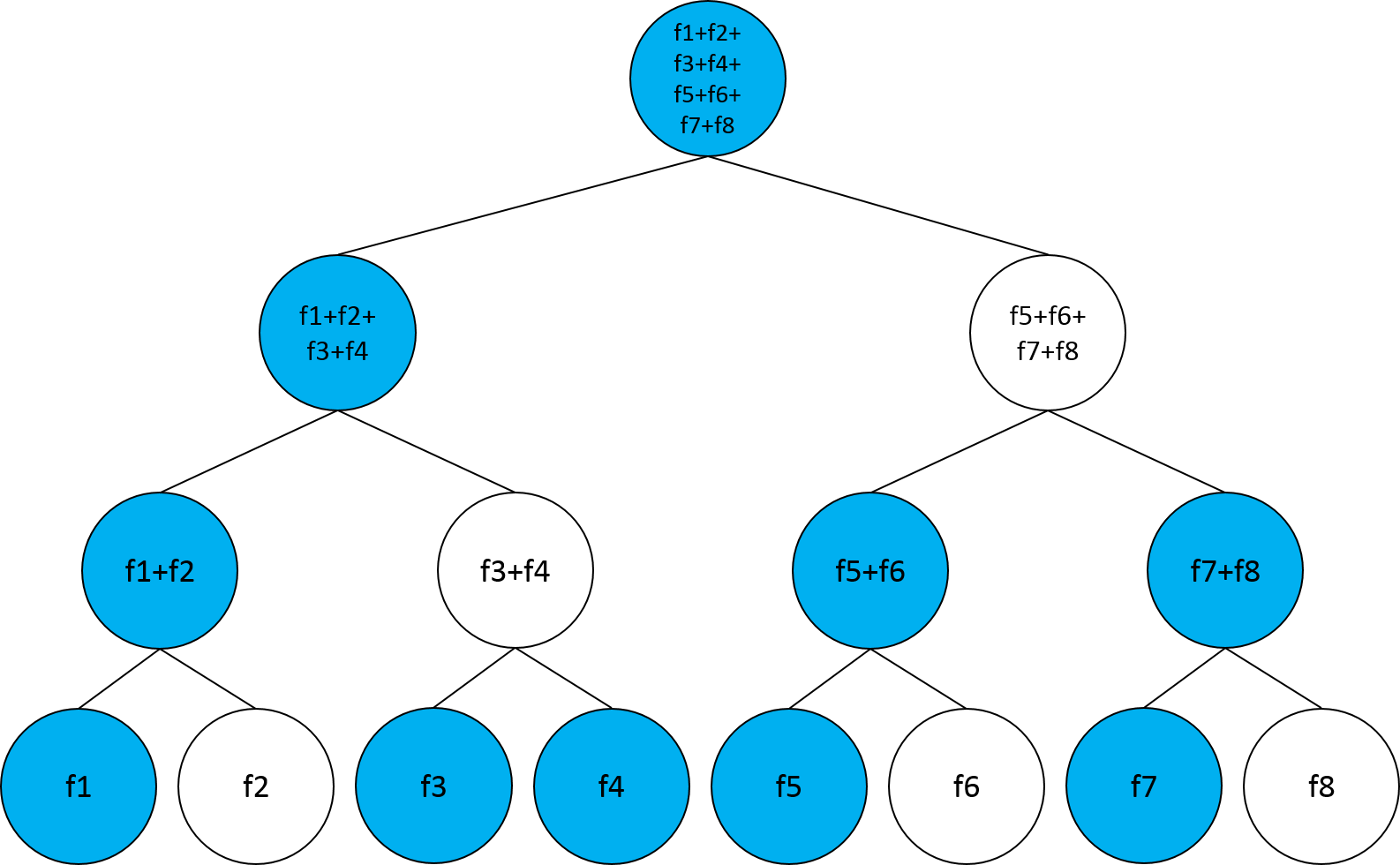}}
\caption{Example of a decodable Treeplication subset with more than $\kfrag=8$ vertices. Out of the $10$ vertices in the subset, $\kfrag=8$ are sufficient to decode the full data unit.}
\label{fig:decsubset2}
\end{figure}

\section{Decodability with Uniform Selection}\label{sec:uniform}
Treeplication is intended for use in a fully distributed storage system where nodes decide in a decentralized way which fragments to store. In the simplest model for decentralized fragment selection, $\nvert$ code fragments are each chosen uniformly and independently (with replacement, so multiplicity is possible) from the $2\kfrag-1$ tree vertices. To accommodate multiplicities in the drawing of vertices, we next define Treeplication {\em multisets}. 
\begin{defn}
A \textbf{Treeplication multiset} for $\tre_{\ell}$ is defined as a multiset of vertices from $\tre_{\ell}$. 
\end{defn}
Every Treeplication multiset can be mapped to a Treeplication subset by removing vertex multiplicities, and for the sake of decodability, it is sufficient to look at this subset. We now derive the probability of obtaining a decodable subset once the above-mentioned selection is performed. We first count the number of decodable subsets given $\kfrag+j$ {\em unique} vertices in the multiset, for $j=0,\ldots,\kfrag-1$; subsequently, we count $\nvert$-multisets mapping to decodable $\kfrag+j$ unique vertices.

For a tree with $d$ layers, define $D_{d,j}$ to be the number of decodable subsets with $2^{d-1}+j$ (unique) vertices. Note that $D_{d,j}=0$ if $j<0$ or $j>2^{d-1}-1$. We partition the decodable subsets to $D_{d,j} = t_{d,j} + r_{d,j}$, where $t_{d,j}$ is the number of subsets that include the root vertex {\em and} are non-decodable without it, while $r_{d,j}$ is the number of all other decodable subsets.

\begin{lemma}\label{lem:count_r}
For a tree with $d$ layers and a Treeplication subset with $2^{d-1}+j$ (unique) vertices, the following recursive relation applies to $r_{d,j}$
\begin{equation}
r_{d,j} = \sum^{j}_{l=0} D_{d-1,l}D_{d-1,j-l} + \sum^{j}_{l=0} D_{d-1,l}D_{d-1,j-l-1}.\label{eq:count_r}
\end{equation}

\end{lemma}

\begin{IEEEproof}
By the definition of $r_{d,j}$, the counted decodable subsets either do not have the root vertex or they are decodable without it. In either case the immediate subtrees of the root must both be decodable themselves. The first and second terms in~\eqref{eq:count_r}, respectively, count these decodable subsets without and with the root vertex in them.
\end{IEEEproof}
The more interesting is the term $t_{d,j}$, which we count next.
\begin{lemma}\label{lem:count_t}
For a tree with $d$ layers and a Treeplication subset with $2^{d-1}+j$ unique vertices, the following recursive relation applies to $t_{d,j}$

\begin{equation}
t_{d,j}  =  2\sum^{j}_{l=0} D_{d-1,l}t_{d-1,j-l},~d>1,j\geq 0;~ t_{1,0} =  1. \label{eq:count_t}
\end{equation}

\end{lemma}

\begin{IEEEproof}
When $d=1$ the tree is just the root vertex, and the root forms a decodable subset that is non-decodable without it (trivially); hence $t_{1,0}=1$. By Lemma~\ref{lem:subtree_decodable}, one immediate subtree of $\tre_d$ must be decodable internally, and by definition of $t_{d,j}$ the other subtree must {\em not} be decodable internally. The former gives the term $D_{d-1,l}$ in~\eqref{eq:count_t} and the latter gives $t_{d-1,j-l}$. To understand why the latter is correct, observe that every decodable subset of $\tre_{d-1}$ that contains its root can be mapped to a decodable subset of $\tre_d$ by replacing the root of $\tre_{d-1}$ by the root of $\tre_d$, assuming that the other subtree $\tre'_{d-1}$ is decodable internally. Also, with this root swap it is clear that $\tre_{d-1}$ is non-decodable without its root if and only if $\tre_d$ is non-decodable without its root. Finally, the factor 2 in~\eqref{eq:count_t} counts the two options to choose the internally-decodable subtree among the left and right subtrees.
\end{IEEEproof}

Once we have an efficient way to count decodable subsets, it is simple to derive the probability to get a decodable subset under independent uniform selection of each of the $\nvert$ code fragments.
\begin{theorem}
For a tree with $d$ layers and a $\nvert$-multiset, each of whose elements is chosen uniformly and independently from the $2^{d}-1$ vertices of $\tre_d$, the probability to get a decodable Treeplication subset is
\begin{equation}
P_{d} = \sum_{j=0}^{\nvert - 2^{d-1}}\frac{D_{d,j}S(\nvert,2^{d-1}+j)(2^{d-1}+j)!}{(2^d-1)^{\nvert}},
\end{equation}
where $S(a,b)$ is the number of ways to partition a set of $a$ elements into $b$ nonempty subsets (also known as the Stirling number of the second kind).
\end{theorem}
\begin{IEEEproof}
Each decodable subset counted by a $D_{d,j}$ can be chosen in $S(\nvert,2^{d-1}+j)(2^{d-1}+j)!$ different ways by the uniform selection. The probability is obtained by normalizing by the total number of choices, decodable or not.
\end{IEEEproof}
We compare the Treeplication scheme under uniform independent selection to standard (uncoded) replication with the same selection policy. For the same input block size $\kfrag$, in replication each choice is one of $\kfrag$ data fragments, while in Treeplication it is one of $2\kfrag-1$ code fragments. The comparison results can be seen in the two middle columns of Table~\ref{tab:uniform_vs_rep} below. The results show the advantage of Treeplication: to get to the same decoding-success\footnote{In replication, decoding success is when every data fragment is selected at least once.} probability of $0.9$, replication needs between 15\%-30\% higher $\nvert$ than Treeplication, which means a higher storage cost for the same availability performance.

\section{Non-uniform Selection}\label{sec:nonuniform}
The uniform selection assumption considered above may not render the optimal tree vertex selection, hence better results may be obtained. With this in mind, in order to improve decodability we now extend the analysis to non-uniform selection. In the non-uniform setup we have $\nvert=\sum_{i=1}^{d}\nvert_i$, where $\nvert_i$ code fragments are chosen (with replacement) from layer $i$ of the tree. Within each layer the selection is as before: each of the $\nvert_i$ code fragments is chosen uniformly and independently from the $2^{d-i}$ vertices of layer $i$. In our analysis we map each $\nvert_i$ to $p_i=1-(1-\frac{1}{2^{d-i}})^{\nvert_i}$, where $p_i$ is the probability that a certain vertex in layer $i$ is selected to the multiset at least once (same for all vertices in the layer). Note that $p_i$ is monotonically increasing with $\nvert_i$, and $p_i=0$ when $\nvert_i=0$. It will be simpler for us to assume that a vertex in layer $i$ is included in the multiset (at least once) with probability $p_i$, independently of the other vertices in the layer, although this assumption is not consistent with the specified discrete parameters $\{n_i\}_{i=1}^{d}$. This assumption is a reasonable approximation when $\nvert_i$ is of the same order as $2^{d-i}$, as required to get decodability with high probability\footnote{For verification we compared the i.i.d model with uniform distribution to the true uniform results of Section~\ref{sec:uniform}, and got almost the same results.}.  The following theorem gives an expression for the decoding probability with non-uniform selection.

\begin{theorem}\label{th:dec_succ_nonuniform}
For a tree $\tre_d$ whose vertices are chosen with probability $p_i$ in layer $i$, the probability to get a decodable Treeplication subset is

\begin{equation}
Q_d = Q_{d-1}^2+2^{d-1} p_d \prod\limits_{i=1}^{d-1}[(1-p_{i})Q_i],~d>1;~Q_1 = p_1. \label{eq:nonuniform_decod}\\
\end{equation}

\end{theorem}

\begin{IEEEproof}
When $d=1$ the tree is just the root vertex, and the subset is decodable if it contains the root vertex, happening with probability $p_1$. For $i=1,\ldots,d-1$, denote by $B_i$ the probability that the subset elements in $T_i$ can decode the leaves of $T_i$ if and only if its root vertex is provided to the subset externally. Then $Q_d=Q_{d-1}^2+2Q_{d-1}p_dB_{d-1}$, because, similar to Lemmas~\ref{lem:count_r},\ref{lem:count_t}, the subset is decodable if both its subtrees are decodable, or if one subtree is decodable, the root is present, and the other subtree is decodable if and only if its root is provided externally. The ``only if'' is required to not count in the second term probabilities already included in the term $Q_{d-1}^2$; the ``if'' part guarantees that the other subtree is decodable when the parent root is present and the other subtree is decodable. $B_{d-1}$ can be calculated with the recursive expression $B_i=2(1-p_i)Q_{i-1}B_{i-1}$, and the initial value $B_1=1-p_1$. Expanding this expression to $B_{d-1}$ and substituting in the previous equation gives~\eqref{eq:nonuniform_decod}.
\end{IEEEproof}
By calculating efficiently $Q_d$ for every {\em selection distribution} $\{\nvert_i\}_{i=1}^{d}$, Theorem~\ref{th:dec_succ_nonuniform} is a useful tool to design non-uniform Treeplication allocations that, for any given $\nvert$, maximize $Q_d$ among all $\{\nvert_i\}_{i=1}^{d}:\sum_{i=1}^{d}\nvert_i=\nvert$. To find the optimal $Q_d$ efficiently, we first prove some properties of optimal selection distributions that significantly reduce the search complexity.

\begin{lemma}\label{lem:pi_monotone}
Every optimal selection distribution satisfies $p_i \leq p_{i-1}$, $\forall i \in [2,d]$.
\end{lemma}
\begin{IEEEproof}
Assume that $p_1,\ldots,p_{i-2}$ are set, and by contradiction that $p_i>p_{i-1}$. From~\eqref{eq:nonuniform_decod} we have
\begin{equation}
Q_i = Q_{i-1}^2+2^{i-1} p_i \prod\limits_{j=1}^{i-1}[(1-p_{j})Q_j],\label{eq:recur_Qi}
\end{equation}
\begin{equation}
Q_{i-1} = Q_{i-2}^2+2^{i-2} p_{i-1} \prod\limits_{j=1}^{i-2}[(1-p_{j})Q_j].\label{eq:recur_Qi1}
\end{equation}
Denote $a:=Q_{i-2}^2$ and $b:= 2^{i-2} \prod\limits_{j=1}^{i-2}[(1-p_{j})Q_j]$. By substituting~\eqref{eq:recur_Qi1} and $a,b$ into~\eqref{eq:recur_Qi}, we get
\begin{equation}
Q_i = (a+bp_{i-1})(a+bp_{i-1}+2bp_i-2bp_{i-1}p_i).\label{eq:diff_Qi}
\end{equation}
Since $a,b$ are independent of $p_{i}$ and $p_{i-1}$, we can see that exchanging between $p_{i}$ and $p_{i-1}$ in~\eqref{eq:diff_Qi} results in an increase in $Q_i$ because
\begin{equation}
(a+bp_{i-1})(a+bp_{i-1}+2bp_i-2bp_{i-1}p_i)  < (a+bp_{i})(a+bp_{i}+2bp_{i-1}-2bp_{i-1}p_i)
\end{equation}
for any $0\leq p_{i-1}<p_{i}\leq 1$. This is a contradiction.
\end{IEEEproof}
The monotonicity of $p_i$ in $i$ implies the following lemma.
\begin{lemma}\label{lem:ni_monotone}
Every optimal selection distribution satisfies $n_{i-1} \geq 2n_{i}$, $\forall i \in [2,d]$.

\end{lemma}
\begin{IEEEproof}
From Lemma~\ref{lem:pi_monotone} we have $1-p_i\geq 1-p_{i-1}$, and from monotonicity of the log function $\log(1-p_i)\geq \log(1-p_{i-1})$.  Thus substituting the definition of $p_i$,$p_{i-1}$ gives that
\begin{equation} \frac{\nvert_{i-1}}{\nvert_{i}} \geq \frac{\log\left(1-\frac{1}{2^{d-i}}\right)}{\log\left(1-\frac{1}{2^{d-i+1}}\right)}\geq 2.\end{equation}
\end{IEEEproof}
With Lemma~\ref{lem:ni_monotone} we can prove the following useful property of optimal selection distributions.
\begin{proposition}\label{prop:ni_squash}
Every optimal selection distribution satisfies $\nvert_{i} \geq \sum_{j=i+1}^{d}\nvert_{j}$, $\forall i \in [1,d-1]$.
\end{proposition}
\begin{IEEEproof}
By induction starting from $i=d-1$. True for base case $i=d-1$ because $\nvert_{d-1}\geq 2\nvert_{d} \geq \nvert_{d}=\sum_{j=d}^{d}\nvert_{j}$, where the first inequality is from Lemma~\ref{lem:ni_monotone}. Assume true for $i$, then showing for $i-1$
\[\nvert_{i-1}\geq 2\nvert_{i} \geq \nvert_i+\sum_{j=i+1}^{d}\nvert_{j}=\sum_{j=i}^{d}\nvert_{j},\]
where the first inequality is from Lemma~\ref{lem:ni_monotone} and the second from the induction hypothesis.
\end{IEEEproof}
Proposition~\ref{prop:ni_squash} is the basis to Algorithm~\ref{alg:search_nonuni} that finds the optimal selection distribution based on searching the small subset of distributions that satisfy the above optimality conditions. In the algorithm we denote by $\{\nvert_i\}_{i=1}^{\ell}$ a selection distribution $\nvert_1,\ldots,\nvert_{\ell},0,\ldots,0$, where the last $d-\ell$ elements of the distribution are $0$. For any selection distribution $S$ we denote by $Q_d(S)$ the result of~\eqref{eq:nonuniform_decod} with $p_i$ corresponding to the $\nvert_i$ of $S$. In the algorithm, $\Qm$ holds the maximum decoding probability among all selection distributions explored so far.

\begin{algorithm}
\caption{Find optimal non-uniform selection distribution}
\begin{algorithmic}[1]
\Function{Search}{$\nvert, d$}
\State $\Qm:=0$
\State Distribute$(1,\nvert,\{\})$

\State \Return $\Qm$

\EndFunction

\Function{Distribute}{$j,B,\{\nvert_i\}_{i=1}^{j-1}$}
	\If{$B==0~||~j>d$}
        \If{$Q_d(\{\nvert_i\}_{i=1}^{j-1})>\Qm$}
		  \State {$\Qm:=Q_d(\{\nvert_i\}_{i=1}^{j-1})$}
	   \EndIf
	   \State \Return
	\EndIf
        \For {$b \in [0,\lfloor B/2 \rfloor]$}
		\State { $\nvert_j := B-b$ }
		\State Distribute$(j+1,b,\{\nvert_i\}_{i=1}^{j})$
	\EndFor

\EndFunction
\end{algorithmic}
\label{alg:search_nonuni}
\end{algorithm}

Thanks to the factor $1/2$ in the for loop of Algorithm~\ref{alg:search_nonuni}, its running time is significantly reduced compared to trivial search. While trivial search needs to explore all compositions of $\nvert$ into up to $d$ sets, only {\em non-squashing} partitions~\cite{Sloan:03} of $\nvert$ are explored by Algorithm~\ref{alg:search_nonuni}. For example when $d=6$,$\nvert=128$, trivial search requires 275584033 steps while Algorithm~\ref{alg:search_nonuni} only 25509.

It turns out that optimal Treeplication codes greatly outperform both replication and uniform Treeplication. The right column of Table~\ref{tab:uniform_vs_rep} shows that compared to optimal Treeplication, the storage cost of replication is higher by 60\% or more for $\kfrag\in\{8,16,32\}$. This is close to quadruple the advantage of uniform Treeplication. An attractive property of the optimal Treeplication distributions is that they have a vast majority of systematic data fragments, which means that the system behaves very similarly to an uncoded replication system, only with a much better full-recovery performance. For example, in the third row of Table~\ref{tab:uniform_vs_rep} the optimal distribution for $\nvert=20$ is $\nvert_1=16,\nvert_2=2,\nvert_3=1,\nvert_4=1$, namely, $4/5$ of the nodes have data fragments. 
\begin{table}[htbp]
\caption{Replication vs. Treeplication (uniform and non-uniform): minimum number of stored fragments $\nvert$ required for decoding probability of 0.9.}
\begin{center}
\begin{tabular}{ |c||c|c|c| }
\hline
 $\kfrag$ & Replication & Treeplication (uniform)  &  Treeplication (non-uniform) \\
 \hline
 2& 5 & 4 &  3\\
 4& 13 & 10  &  8\\
 8& 33 & 26 &   20\\
 16& 79 & 66 &   49\\
 32& 181 & 157 &  113\\
 \hline
\end{tabular}
\label{tab:uniform_vs_rep}
\end{center}
\end{table}
Further results comparing the three schemes are given in Fig.~\ref{fig:dcomp} where the decoding probability is plotted as a function of $\nvert$ for $\kfrag=8,16$.

\begin{figure}
    \centering
    \begin{subfigure}[b]{0.45\textwidth}
        \includegraphics[width=\textwidth]{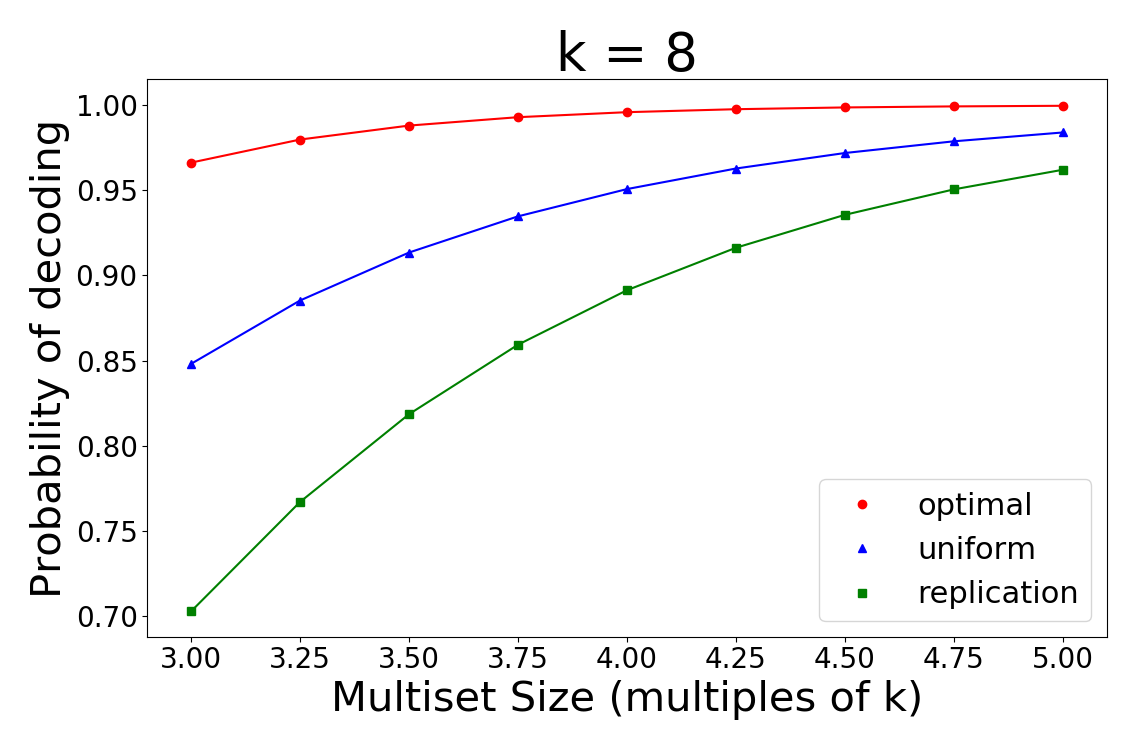}
        \label{fig:k8}
    \end{subfigure}
    ~
    \begin{subfigure}[b]{0.45\textwidth}
        \includegraphics[width=\textwidth]{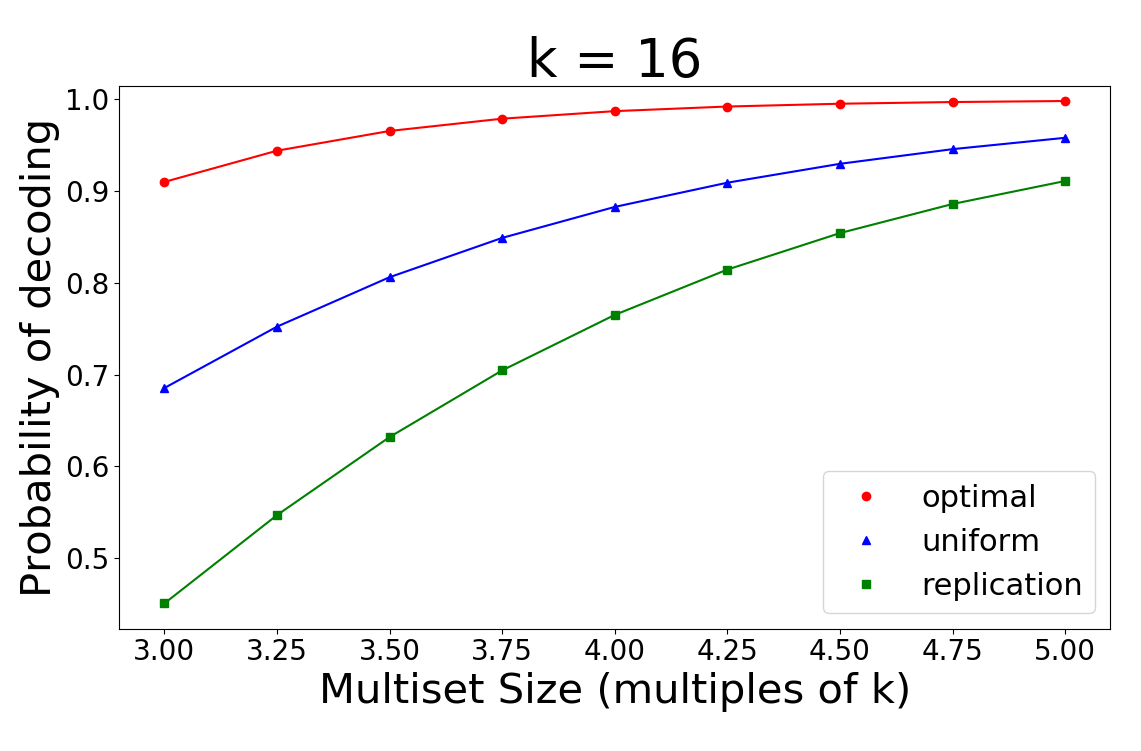}
        \label{fig:k16}
    \end{subfigure}

    \caption{Probability of obtaining a decodable subset as a function of the multiset size $\nvert$ given in multiples of $\kfrag$, for $\kfrag=8,16$. Three curves compare: optimal Treeplication, uniform Treeplication, and replication.}\label{fig:dcomp}
\end{figure}
}

\itw{
\section{Fragment recovery and communication cost}\label{RC}
In the successful case of having a decodable subset of $\tre_d$, the distributed storage system needs to have the $\kfrag$ data fragments recovered by the nodes storing the decodable subset. The recovery process is done in a distributed way, where each data fragment is recovered by one node storing a code fragment, using code fragments from other nodes if necessary. For any subset size $\nvert\geq \kfrag$, $\kfrag$ nodes are chosen to each recover a different data fragment, in a way that communication cost is minimized. Data fragments that appear as leaves (systematic code fragments) in the subset are trivially recovered locally with no communication; the remaining data fragments are recovered by non-leaf vertices that receive code fragments (both systematic and not) of other vertices to recover the assigned data fragment. The cost in terms of communications required for this recovery is the {\em total} number of code fragments communicated to recover all $\kfrag$ data fragments, and it should be minimal.

\subsection{Minimal-communication recovery algorithm}\label{subsec:recovery_alg}
This sub-section presents Algorithm~\ref{alg:recover_frags}, which finds the minimal-communication recovery and counts the number of fragment transmissions. At the start of the algorithm we have a decodable subset (with $\kfrag$ or more fragments) mapped to a tree. Each fragment (tree vertex) is stored by a node in the system, and all nodes know the vertices in the subset. Algorithm~\ref{alg:recover_frags} is run by each of these nodes, to determine which data fragment (leaf vertex) to recover (if any), and which fragments (tree vertices) to request from the other nodes in the subset. Before presenting the algorithm, we prove properties regarding node selection for minimal-communication recovery. In the sequel, a {\em present} resp. {\em missing} vertex is a vertex that is {\em in} resp. {\em not in} the decodable subset. We define a {\em missing-vertex path} as a path in the tree in which all vertices are missing.

\begin{lemma}\label{lem:no_two_paths}
If a Treeplication subset is decodable, then 1) there is no missing-vertex path between a leaf and the root, and 2) no vertex (present or missing) has missing-vertex paths connecting its two children with two leaves.
\end{lemma}
\begin{IEEEproof}
The existence of a missing-vertex path from leaf to root contradicts decodability because in that case no present code fragment depends on that leaf. Two missing-vertex paths ending at vertices with a common parent vertex $x$ imply that both subtrees directly under $x$ are non-decodable (by condition 1 above), thus violating Lemma~\ref{lem:subtree_decodable}.
\end{IEEEproof}

\begin{proposition}\label{prop:recover_vert}
Suppose present vertex $x$ recovers leaf vertex $y$ if and only if there is a missing-vertex path between a child of $x$ and $y$. Then in a decodable Treeplication subset, each missing leaf is recovered by a single unique vertex, and this vertex is the lowest one capable of recovering $y$.
\end{proposition}
\begin{IEEEproof}
From condition 1 of Lemma~\ref{lem:no_two_paths}, each missing leaf must have a path upward ending at a present vertex; from condition 2 there cannot be another leaf that is in missing-vertex path ending at a child of $x$. These prove that every leaf will be recovered by a unique present vertex. $x$ is the lowest present vertex in the tree that can recover $y$, because it is at the end of a missing-vertex path from $y$, making it the lowest vertex whose code fragment has $y$ as argument.
\end{IEEEproof}

Building on Proposition~\ref{prop:recover_vert}, Algorithm~\ref{alg:recover_frags} now finds the vertices recovering the missing data fragments with minimal communication (the present data fragments are recovered locally with no communication, and are not handled by Algorithm~\ref{alg:recover_frags}). Each of the vertices chosen for recovery is the lowest possible in the tree able to recover the corresponding data fragment, hence requires the least amount of communication. A vertex evaluated to ``false'' in line 9 is not recovering any data fragment, and can be discarded as redundant (this happens when the decodable subset has more than $k$ vertices). The variable $\mathrm{sum}$ holds the aggregate number of code fragments communicated to the nodes recovering the data fragments. The explicit identities of the communicated code fragments can be extracted from the identities of the vertices reached in line 6. If Algorithm~\ref{alg:recover_frags} terminates without recovering all missing data fragments, from Proposition~\ref{prop:recover_vert} we know that the subset is not decodable.
\begin{algorithm}
\caption{Fragment recovery}
\begin{algorithmic}[1]
\State $\mathrm{sum} := 0$
\For{each present vertex $x$}
	\State $\mathrm{sum}_{x} := 0$
	\For{each path downwards from $x$}
		\State { traverse path until a present vertex or a missing leaf reached}
		\If{a vertex that is present reached} $\mathrm{sum}_{x}$++ \EndIf
	\EndFor
	\If{missing leaf was reached in a downward path} $\mathrm{sum}=\mathrm{sum}+\mathrm{sum}_{x}$ \hspace{.1in} // $x$ is recovering a leaf \EndIf
\EndFor

\State return $\mathrm{sum}$

\end{algorithmic}
\label{alg:recover_frags}
\end{algorithm}

\begin{figure}[htbp]
\centerline{\includegraphics[width=250pt]{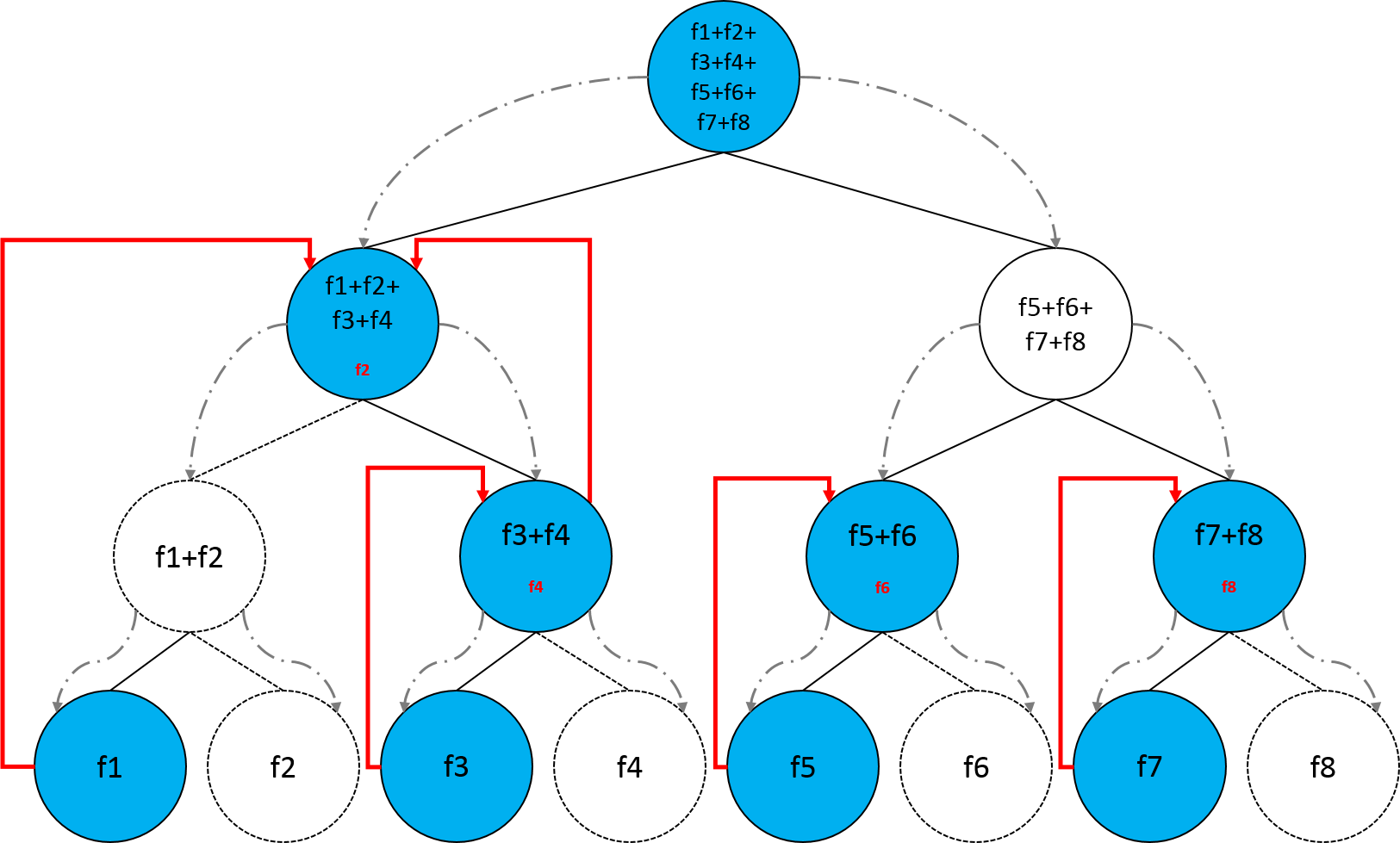}}
\caption{Run of Algorithm~\ref{alg:recover_frags} on a decodable Treeplication subset with $9$ distinct code fragments.}
\label{fig:recovery}
\end{figure}
}

Fig.~\ref{fig:recovery} illustrates an example of a run of Algorithm~\ref{alg:recover_frags}. A red label $f_j$ in vertex $x$ shows the algorithm's finding in line 9 that the node storing $x$ is to recover the missing data fragment $f_j$. A solid red arrow from vertex $z$ to vertex $x$ represents the algorithm's finding in line 6 that the present code fragment $z$ is needed by the node storing $x$ to recover its assigned data fragment. The $\mathrm{sum}$ output of Algorithm~\ref{alg:recover_frags} is the number of red arrows in Fig.~\ref{fig:recovery}, which is the total communication cost to be incurred on recovering all data fragments: $5$ code fragments (4 systematic and 1 not) are communicated to recover all $8$ data fragments. 

It is important to note that our assumption that the input to Algorithm~\ref{alg:recover_frags} is a decodable subset is only for simplicity of presentation. By a small change to the algorithm, we can detect a non-decodable subset (and return $\mathrm{sum}=\infty$) when the condition in line 9 evaluates to ``true'' fewer times than the number of missing leaves. 
    
The following proposition gives the {\em worst case} communication cost of Treeplication.

\begin{proposition}
Recovering a data unit from a decodable Treeplication subset requires communicating at most 
$\kfrag-1$ code fragments in total. 
\end{proposition}
\begin{IEEEproof}
We first prove that without loss of generality, the decodable subset has exactly $\kfrag$ present vertices. If not, we can remove the $x$ vertices that evaluate to ``false'' in line 9 of Algorithm~\ref{alg:recover_frags}, and remain with only present leaves and vertices that each recovers a unique missing leaf; these together sum to $\kfrag$. Now we observe in lines 5,6 of Algorithm~\ref{alg:recover_frags} that the communication count is incremented only for the {\em first} present vertex reached in a path downward from $x$. So looking upward from a present vertex, it can only be communicated to one vertex, which is the first present vertex in its path to the root. This shows at most $\kfrag$ communicated fragments. To show upper bound of $\kfrag-1$, we observe that in any subset there is a present vertex that has no present vertices above it in the tree (e.g. the root). This vertex is (or, if plural, these vertices are) not communicated during recovery, thus bounding the total communication at $\kfrag-1$ or less. 
\end{IEEEproof}
Note that the bound of $\kfrag-1$ is worst-case tight, as the subset in Fig.~\ref{fig:decsubset} in fact requires $\kfrag-1=7$ fragments to be communicated in recovery. The nice thing about the number $\kfrag-1$ is that it equals the number of code fragments a node needs for {\em centralized full recovery} with MDS codes, meaning that Treeplication supports distributed full recovery with no extra cost. Moreover, in the following we see that on average the communication cost is significantly lower than this worst case.   

\itw{
To evaluate the {\em average} recovery communication cost of Treeplication, we first show in Table~\ref{table:2} the empirical average number of code fragments communicated per instance of a decodable subset. Treeplication is implemented using the optimal (non-uniform) selection parameters $\{n_i\}_{i=1}^{d}$ found by Algorithm~\ref{alg:search_nonuni}, and uniform sampling (with replacement) of $n_i$ vertices in layer $i$. The minimal-communication recovery per simulation instance is found using Algorithm~\ref{alg:recover_frags}.  We compare the communication cost to similar decentralized recovery using systematic MDS codes with block length $2\kfrag-1$ (identical to the vertex count of Treeplication) also simulated as a uniform and independent selection (of $\nvert$ fragments from the $2\kfrag-1$ code symbols). For both schemes we used the same $\nvert=3\kfrag$, which is a common replication factor in pure replication settings such as the default in~\cite{Ghemawat}. We can see that Treeplication is very economical in communication, requiring very small numbers of fragments per instance (in particular much smaller than the worst-case $\kfrag-1$). When using MDS codes, recovery of non-systematic fragments requires a heavy load of $\kfrag-1$ fragments per recovering node, which results in an order of magnitude or more higher communication cost per instance.

\begin{table}[htbp]
\caption{Treeplication vs. MDS: communication cost (n=3k).}
\begin{center}
\begin{tabular}{ |c||c|c| }
\hline
 $\kfrag$  & \# fragments Treeplication & \# fragments MDS \\
 \hline
 4& 0.35  & 1.82\\
 8& 1.18 &  10.64\\
 16& 2.88 &  49.62\\
 32& 6.552 &  213.10\\
 \hline
\end{tabular}
\label{table:2}
\end{center}
\end{table}

}
Next we derive analytic expressions for the expected communication cost of Treeplication.
\subsection{Derivation of the expected communication cost}
Throughout the forthcoming analysis, we assume decoding subsets are sampled according to the non-uniform selection of Section~\ref{sec:nonuniform}, that is, a tree vertex in layer $i$ is present in the subset with probability $p_i$, independently from other vertices. We also use the notation $Q_i$ from Section~\ref{sec:nonuniform} to denote the probability that a tree with $i$ layers is decodable under this sampling (recall the efficient calculation of $Q_i$ by Theorem~\ref{th:dec_succ_nonuniform}). The objective of this analysis is to calculate the expected total communication cost to recover all fragments of a data unit, and the expectation is over all {\em decodable} subsets (we exclude from the analysis sampling instances that result in non-decodable subsets; in fact, this necessary exclusion makes the analysis more challenging.) 
The cost we analyze throughout is that of Algorithm~\ref{alg:recover_frags}, which is the minimal for every instance. Following the terminology of Section~\ref{subsec:recovery_alg}, each data fragment is recovered by a unique present vertex, and to each of the $\kfrag$ recovering vertices, zero or more code fragments are communicated. Thanks to the symmetry of data fragments, it is useful to define the expected cost to recover {\em a single} data fragment, and obtain the expected {\em total} cost as $\kfrag$ times this number. We get the expectation of the communication cost per data fragment by deriving the full distribution of the communication cost, defined next.  
\begin{defn}
Let $C_d(N)$ be the probability that in a decodable subset of $\tre_{d}$ a particular data fragment is recovered with $N$ communicated code fragments. 
\end{defn}
We have $\sum_{N\geq 0}C_d(N)=1$, and from symmetry $C_d(N)$ is the same for any data fragment. Since we are only interested in analyzing the cost of {\em decodable} subsets (we assume that for non-decodable subsets the algorithm will halt and not invoke any communications), we define $C_d(N)$ as a probability {\em conditioned on decodability}. The next definition is similar to $C_d(N)$, only defining the {\em joint} probability.
\begin{defn}\label{def:cost_dec_joint}
Let $P_i(N)$ be the probability that a subset of $\tre_{i}$ is decodable, and a particular data fragment is recovered with $N$ communicated code fragments. 
\end{defn}
We changed the tree index from $d$ to $i$ in Definition~\ref{def:cost_dec_joint}, because we will make a recursive use of $P_i(N)$.
The following is a similar definition, only specific to recovery by the root. 
\begin{defn}\label{defn:root_path}
Let $A_i(N)$ be the probability that a subset of $\tre_{i}$ is decodable, and a particular data fragment is recovered with $N$ communicated code fragments, given that the root is present and that there is a missing-vertex path from one of its children to the leaf of the recovered data fragment. 
\end{defn}

In the language of Proposition~\ref{prop:recover_vert}, Definition~\ref{defn:root_path} enumerates the communication cost in cases where the root is chosen to recover the particular data fragment. One final definition is needed to carry out the recursive calculation of the communication-cost distribution. 
\yc{
\begin{defn}\label{defn:traversal}
Let $F_i(N)$ be the probability that a subset of $\tre_{i}$ is decodable, and has $N$ present vertices with no present ancestors.
\end{defn}
Note that in particular $N=1$ in subsets where the root is present. Definition~\ref{defn:traversal} is useful because it will help capture the fragment count $\mathrm{sum}_{x}$, incremented in line 6 of Algorithm~\ref{alg:recover_frags} every time a present vertex is reached in the traversal downward from $x$. We now give a series of lemmas that together facilitate the recursive calculation of $P_i(N)$, and in turn of $C_d(N)$.    
}
\begin{lemma}\label{lem:leaf_comms_p}
$P_i(N)$ can be calculated by
\begin{equation}\label{eq:leaf_comms_p}
P_i(N) = Q_{i-1}P_{i-1}(N) + A_i(N)p_i\prod_{l=1}^{i-1} (1-p_l)  + p_i  \prod^{i-1}_{l=1}(1-p_l) \sum_{j=1}^{i-1}  \left[ 2^{j-1} P_j(N) \prod^{i-1}_{\ell=1, \ell \neq j} Q_\ell \right], ~ i>1 .
\end{equation}

\begin{equation}\label{eq:prob_comms_i1_N}
P_1(N) =  0, ~ N>0 .
\end{equation}

\begin{equation}\label{eq:prob_comms_i1_0}
P_1(0) =  p_1 .
\end{equation}

\end{lemma}

\begin{IEEEproof}
For~\eqref{eq:prob_comms_i1_N},~\eqref{eq:prob_comms_i1_0} in which $i=1$, the decodability probability is $p_1$, and no communicated fragments are required ($N=0$). When $i>1$, we divide to three mutually exclusive cases, each corresponding to a summand in the right-hand side of~\eqref{eq:leaf_comms_p}. 

\underline{Case 1:} Both subtrees of the root are independently decodable. In that case we distinguish between the subtree that contains the leaf of the recovered data fragment (henceforth called the "recovered leaf") and the other subtree. Then the probability decomposes to a product of $P_{i-1}(N)$ for the recovery in the recovered leaf's subtree, times the probability $Q_{i-1}$ that the other subtree is decodable. In the remaining cases one of the root's subtrees is not independently decodable, which implies that the root recovers a leaf vertex, either the recovered leaf (Case 2) or a different one (Case 3).
 
\underline{Case 2:} The recovered leaf is recovered by the present tree root. Recall from Proposition~\ref{prop:recover_vert} that this case implies a missing-vertex path between the recovered leaf and a child of the root. The probability to have a present root and this missing-vertex path is $p_i\prod_{l=1}^{i-1} (1-p_l)$. The remaining term $A_i(N)$ in the second summand is, by definition, the probability that $N$ code fragments are communicated, given that recovery is by the root.   

\underline{Case 3:} The recovered leaf is recovered by a vertex other than the root, while the present root recovers a different leaf, which we call ``another leaf''. The former condition distinguishes from Case 2, and the latter distinguishes from Case 1 because it implies a root's subtree that is not independently decodable.
Given a decodable subset, every missing-vertex path between a child of the root and another leaf defines a partition of the remaining $2^{i-1}-1-i$ tree vertices (vertices of $\tre_{i}$ not in this path and not the root) to subtrees. Each such partition has one subtree of $\ell$ layers, for each $\ell\in\{i-1,i-2,\ldots,1\}$. According to Lemma~\ref{lem:subtree_decodable}, the root and each of the vertices of the missing-vertex path, except the leaf, must have at least one immediate subtree that is independently decodable. For each of these vertices, the subtree in the direction of the missing-vertex path is not independently decodable by Lemma~\ref{lem:no_two_paths} (part 1). Thus the subtrees in such a partition must all be independently decodable. From the set $\{i-1,i-2,\ldots,1\}$ we take $j$ to be the number of layers of the special decodable subtree containing the recovered leaf. Given $j$, there are $2^{j-1}$ different missing-vertex paths ending in another leaf, each defining a different such partition. The third summand of~\eqref{eq:leaf_comms_p} sums over all possible partitions the probability of recovering the recovered leaf with $N$ fragments, within those partitions. The terms of this summand are: 1) $p_i$ the probability that the root is present, 2) $\prod^{i-1}_{l=1}(1-p_l)$ the probability that the path defining the partition is a missing-vertex path (same probability for all partitions), 3) sum over all partitions, using the size index $j$, of the probability that the recovered leaf is recovered in its subtree with $N$ fragments ($P_j(N)$), and the other subtrees are decodable $\prod^{i-1}_{\ell=1, \ell \neq j} Q_\ell$.

\end{IEEEproof}

Next, we calculate $A_i(N)$ recursively from $A_{i-1}(\cdot)$ and $F_{i-1}(\cdot)$.
\begin{lemma}\label{lem:subtree_comms_top_p}
$A_i(N)$ can be calculated by
\begin{equation}\label{eq:a_igt2}
A_i(N) = \sum\limits_{l=1}^{2^{i-2}}\left[F_{i-1}(l) A_{i-1}(N-l)\right], ~i>2
\end{equation}

\begin{equation}\label{eq:a_i2n1}
A_2(1) = p_{1}, 
\end{equation}

\begin{equation}\label{eq:a_iotherwise}
A_i(N) = 0,~ \mathrm{otherwise}.
\end{equation}

\end{lemma}
\begin{IEEEproof}
When $i=2$, the recovered leaf is a child of the root, hence both ends of the missing-vertex path are the recovered leaf itself. Conditioned on the existence of this missing-vertex path and the present root, the tree is decodable if and only if the other child of the root is present, which occurs with probability $p_1$. In this case $N=1$ fragment is communicated.

For $i>2$, we split the $N$ communicated fragments to $l$ coming from the root's subtree not including the recovered leaf (which we call the former subtree), and $N-l$ from the subtree of the recovered leaf (which we call the latter subtree). We know that the latter subtree is not independently decodable (from the existence of the missing-vertex path), so from Lemma~\ref{lem:subtree_decodable} the former subtree must be independently decodable. Having $l$ fragments communicated from the former subtree which is independently decodable occurs with probability $F_{i-1}(l)$ by definition. These $l$ fragments together with the root fragment can recover the root of the latter subtree; requiring $N-l$ fragments from this subtree occurs with probability $A_{i-1}(N-l)$ because the conditions in the definition of $A_{\ell}(\cdot)$ are satisfied for the latter subtree when its root is known.

Summing over all possible values of $l$, we get~\eqref{eq:a_igt2}.

\end{IEEEproof}
Finally, the next lemma shows how to calculate $F_{i}(N)$ recursively from $F_{i-1}(\cdot)$. 
\begin{lemma}\label{lem:subtree_comms_p}
$F_i(N)$ can be calculated by

\begin{equation}\label{eq:f_ingt1}
F_i(N) = (1-p_i) \sum_{l=1}^{N-1}\left[F_{i-1}(l)F_{i-1}(N-l) \right], ~N>1
\end{equation}

\begin{equation}\label{eq:f_i1}
F_i(1) = p_i \left[Q_{i-1}^2 + 2^{i-1}\prod_{j=1}^{i-1}[(1-p_j)Q_j]\right].
\end{equation}

\end{lemma}
\begin{IEEEproof}
When the root is present, only one vertex (the root itself) has no present ancestors. Hence $N=1$, and the product in~\eqref{eq:f_i1} gives the probability that the root is present and the tree is decodable (note that the right-hand side of~\eqref{eq:f_i1} is similar to~\eqref{eq:nonuniform_decod}, but not identical because here it is required that the root is present). When $N>1$, the root is not present, and the number of present vertices with no present ancestors divides to $l$ in one subtree and $N-l$ in the other. This gives the product $F_{i-1}(l)F_{i-1}(N-l)$ for the probability, and summing over all $l$ and multiplying by the probability $1-p_i$ that the root is not present, we get~\eqref{eq:f_ingt1}.

\end{IEEEproof}

With Lemmas~\ref{lem:subtree_comms_top_p},~\ref{lem:subtree_comms_p} and~\ref{lem:leaf_comms_p}, we can efficiently calculate $P_i(N)$ for any $i$ and $N$, and for any selection distribution $\{p_i\}_{i=1}^{d}$. Then the expected communication cost for decodable subsets of $d$-layer Treeplication is taken simply as 
\begin{equation}E\triangleq 2^{d-1}\sum_{N\geq 0}N\cdot C_d(N) = 2^{d-1} \frac{\sum_{N\geq 0}N\cdot P_d(N)}{Q_d},\label{eq:av_comm_cost}\end{equation}
and the equality follows from the elementary probability-theory relation $P(X|Y)=P(X,Y)/P(Y)$ (where $Y$ is the event that the subset is decodable). 

Evaluating $E$ in~\eqref{eq:av_comm_cost} for $d = \{3, 4, 5, 6\}$ ($\kfrag = \{4, 8, 16, 32\}$) using the optimal non-uniform distribution with $\nvert$, we get the results in Table~\ref{table:2}. For comparison, in each row we include the empirical results from Section~\ref{subsec:recovery_alg}, and observe their good match.   

\begin{table}[htbp]
\caption{Treeplication's expected communication cost ($\nvert = 3\kfrag$).}
\begin{center}
\begin{tabular}{ |c||c|c||c|c|  }
\hline
 $\kfrag$  & \# comm. fragments \\
 \hline
 4& 0.357 (empirical: 0.350)  \\
 8& 1.143 (empirical: 1.180) \\
 16& 2.830 (empirical: 2.880) \\
 32& 6.524 (empirical: 6.552) \\
 \hline
\end{tabular}
\label{table:expectedcomms}
\end{center}
\end{table}
\section{Tree Health}\label{sec:health}
So far in the paper, a Treeplication multiset was evaluated with respect to its instantaneous properties: whether it is decodable, and how efficiently it can recover the data unit. However, in a real distributed storage system, multisets evolve in time due to changes in node availability. Therefore, in this and the next sections we extend the scope to consider {\em forward-looking properties} of the multiset, e.g., its robustness to possible future events of nodes going unavailable. Within that scope, one decodable multiset may be ``better'' than another decodable multiset because it is more likely to remain decodable after some fixed number of nodes go unavailable. We refer to this quality of Treeplication multisets as their {\em tree health}, and develop tools to evaluate and compare multisets with respect to a concrete tree-health measure. Comparing multisets' health is useful in storage systems because it allows to allocate resources efficiently to data units according to health, thereby maximizing the overall system resilience. The tree health is parametrized by an integer $l$, which counts the number of nodes becoming unavailable; $l$ captures the degree of robustness expected from the multiset. Throughout the discussion we assume that the nodes becoming unavailable are drawn uniformly from the multiset. 

The health of a Treeplication multiset depends on its specific present vertices (and multiplicities), and therefore we seek an efficient method to measure the health without exhaustively enumerating all possible combinations of $l$ node losses. To that end, we next prove decodability conditions that facilitate the formulation of a tractable tree-health measure. First we define a subset of the Treeplication tree we call a {\em diagonal}.   

\begin{defn} \label{defn:diagonal}
A set of vertices from $\tre_{d}$ is called a \textbf{diagonal} if it forms a path downward starting at an arbitrary vertex and ending in a leaf vertex.  \end{defn}

A special case of Definition~\ref{defn:diagonal} is a diagonal consisting of a single leaf vertex. A Treeplication tree decomposed to a union of disjoint diagonals is called a {\em diagonal cover}. Next we prove that every Treeplication tree has a (non unique) diagonal cover. 

\begin{proposition} \label{prop:num_diags_cover}
A Treeplication tree $\tre_{d}$ can be decomposed as a diagonal cover with $k$ diagonals, where $1$ diagonal is of size $d$, and for each $i\in\{1,\ldots,d-1\}$ there are $2^{d-i-1}$ diagonals of size $i$. 
\end{proposition}

\begin{figure}[htbp]
\centerline{\includegraphics[width=260pt]{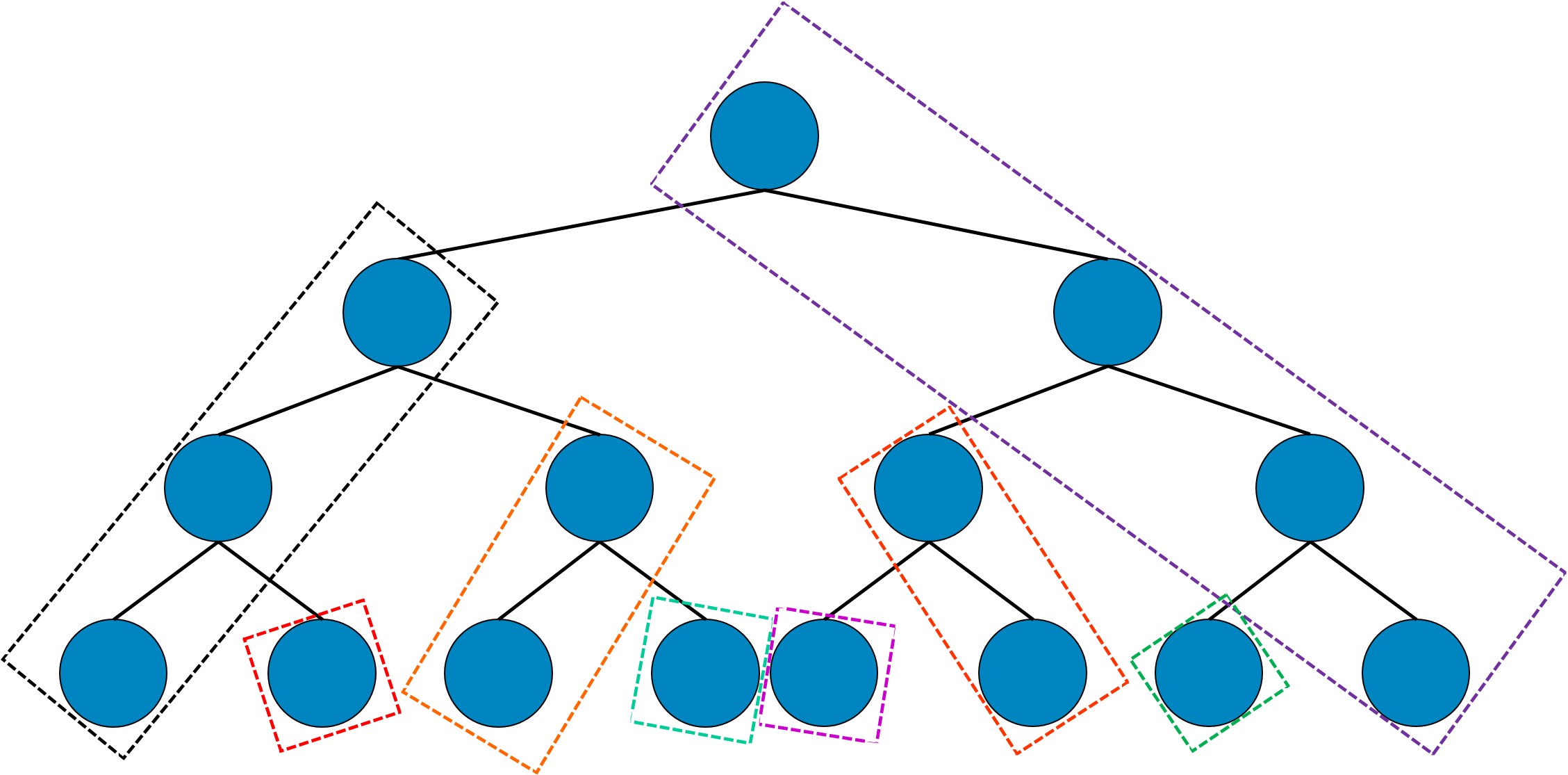}}
\caption{A decomposition of $\tre_{4}$ as a diagonal cover with $8$ diagonals.}
\label{fig:8diagonalcover}
\end{figure}

The proof of Proposition~\ref{prop:num_diags_cover} is immediate by successively assigning diagonals where each diagonal is a path from the top-most vertex not previously assigned to a diagonal, to any leaf below it. Fig.~\ref{fig:8diagonalcover} shows a tree of $4$ layers decomposed into a diagonal cover of $k=8$ diagonals. As stated by Proposition~\ref{prop:num_diags_cover}, one diagonal in the cover is of size $4$, another of size $3$, two of size $2$, and four of size $1$.

Using the decomposition to diagonal covers we can prove the following (sufficient and necessary) decodability condition for Treeplication subsets.

\begin{theorem} \label{th:1vertex_diagonal}
A Treeplication subset for $\tre_d$ is decodable if and only if it has at least one diagonal cover in which every diagonal has at least one present vertex.
\end{theorem}

\begin{IEEEproof}
For sufficiency, we assume there is such a diagonal cover and need to prove decodability. We prove by induction on $d$. For $d=1$, the diagonal cover consists of one leaf, and decodability follows trivially if this leaf is present. For the induction hypothesis, suppose the condition is sufficient for every $\tre_i$ with $i\in\{1,\ldots,d-1\}$. To prove the induction step, we examine the single diagonal of $\tre_d$ with size $d$. Each vertex of this diagonal has under it a subtree where the condition is satisfied. Hence all these subtrees are independently decodable by the induction hypothesis. Now we can show that any present vertex in this diagonal, together with vertices in the independently decodable subtrees, can recover the vertex below it in the diagonal. Applying this iteratively, we can recover the leaf of this diagonal, which is the only leaf not in the independently decodable subtrees.

For necessity, we first construct diagonals by taking each diagonal to include a leaf and all vertices in a path from it upward until the lowest present vertex (if a leaf is present, the diagonal only includes this leaf). If the subset is decodable, then by Lemma~\ref{lem:no_two_paths} (part 2) no two leaves share the same lowest present vertex in their paths upward. This implies that the diagonals are disjoint, and from Lemma~\ref{lem:no_two_paths} (part 1) each diagonal has one present vertex. To complete the diagonal cover, we extend each existing diagonal upward until reaching the top-most vertex not already in a diagonal. This extension results in a diagonal cover, and can only add present vertices to the diagonals, thus the condition is satisfied.
\end{IEEEproof}

Theorem~\ref{th:1vertex_diagonal} provides a sufficient and necessary decodability condition on the diagonals of $\tre_d$'s diagonal covers. By that, it reduces the decodability of the entire subset to the simpler condition that individual diagonals in a cover are non-empty, that is, have at least one present vertex each. One challenge still remaining is that the diagonal covers are not disjoint, and the union probability among all covers to meet this condition cannot be simply calculated as a sum of probabilities for the individual covers (in general the sum of individual probabilities will only give an upper bound on the decodability probability). This challenge motivates defining a tree-health measure that picks one diagonal cover, with respect to which the robustness of the Treeplication multiset is evaluated. The special diagonal cover proposed for this health measure is defined next.


\begin{defn} \label{defn:principal_diagonal_cover}
Given a Treeplication multiset for $\tre_d$, we call a diagonal cover a \textbf{$l$-principal diagonal cover} if its diagonals have maximum average probability of being non-empty after the loss of $l$ multiset elements.
\end{defn}

The motivation to pick out principal diagonal covers from all possible covers is that the principality condition of Definition~\ref{defn:principal_diagonal_cover} makes those covers more likely to fulfill the sufficient condition of Theorem~\ref{th:1vertex_diagonal}. In that sense, we replace the union of covers considered in Theorem~\ref{th:1vertex_diagonal} by one strong candidate that is the $l$-principal diagonal cover. Based on that motivation, we choose the following tree health measure.
\begin{defn}\label{defn:lhealth}
For a Treeplication multiset define the \textbf{principal $l$-health} as the average probability, over the diagonals of a $l$-principal diagonal cover, that the diagonal is non-empty after the loss of $l$ multiset elements.
\end{defn}

Next we give more definitions that will be useful for finding $l$-principal diagonal covers and calculating the principal $l$-healths of multisets.

\begin{defn} \label{defn:vertex_weight}
Given a Treeplication multiset for $\tre_d$ we denote by $v_{i,j}$ the number of times the $j$-th vertex of the $i$-th layer of $\tre_d$ appears in the multiset. We call $v_{i,j}$ the \textbf{vertex weight} of the $i,j$ vertex.
\end{defn}	

For the purpose of Definition~\ref{defn:vertex_weight} we number the tree vertices from left to right in each layer. Additional weight definitions are given next for multisets, diagonals and diagonal covers.

\begin{defn}
Given a Treeplication multiset for $\tre_d$ we define the \textbf{multiset weight} as the sum over all vertices of $\tre_d$ of the vertex weights. 
\end{defn}	  

Note that the multiset weight is simply the number of elements in the multiset, also denoted $\nvert$ in Sections~\ref{sec:uniform},\ref{sec:nonuniform}.

\begin{defn}
For a Treeplication multiset and a diagonal of $\tre_d$ we define the \textbf{diagonal weight} as the total number of times the vertices of this diagonal appear in the multiset.
\end{defn}	

\begin{defn}
For a Treeplication multiset and a diagonal cover of $\tre_d$ we define the \textbf{weight profile} $W \triangleq \{w_j | j \in \{1,\ldots,\kfrag\}\}$, where $w_j$ is the weight of the $j$-th diagonal in the cover.  
\end{defn}	

\begin{figure}[htbp]
\centerline{\includegraphics[width=260pt]{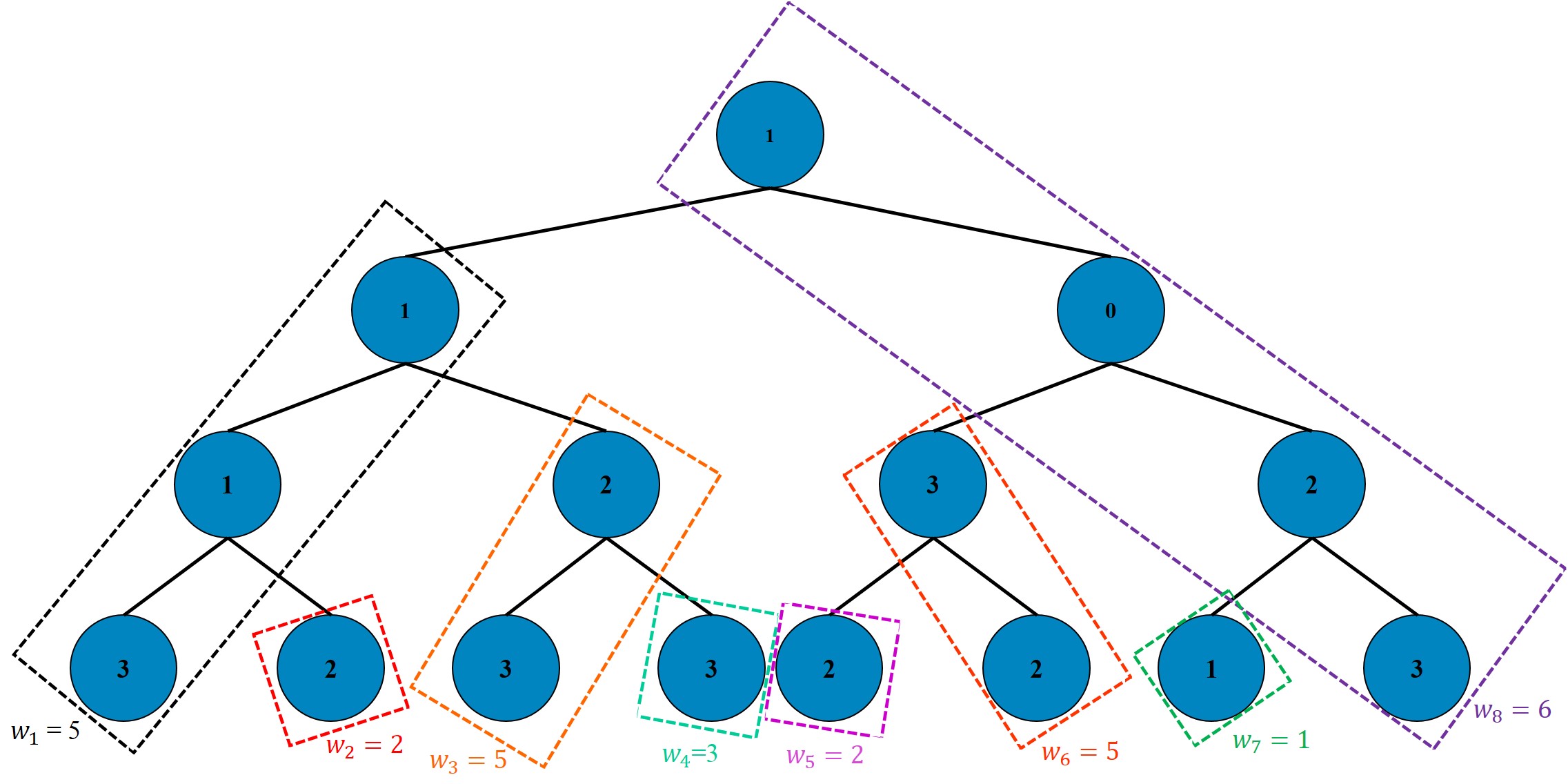}}
\caption{A diagonal cover of a sample Treeplication multiset ($\kfrag=8$). Each vertex is marked by its vertex weight, and each diagonal is marked by its diagonal weight $w_j$.}
\label{fig:8diagonalcover_before}
\end{figure}

Note that for any diagonal cover, the sum $\sum_{j=1}^{\kfrag}w_j$ equals the multiset weight. Fig.~\ref{fig:8diagonalcover_before} shows for the diagonal cover of Fig.~\ref{fig:8diagonalcover} the vertex and diagonal weights of a sample multiset.

\begin{figure}[htbp]
\centerline{\includegraphics[width=260pt]{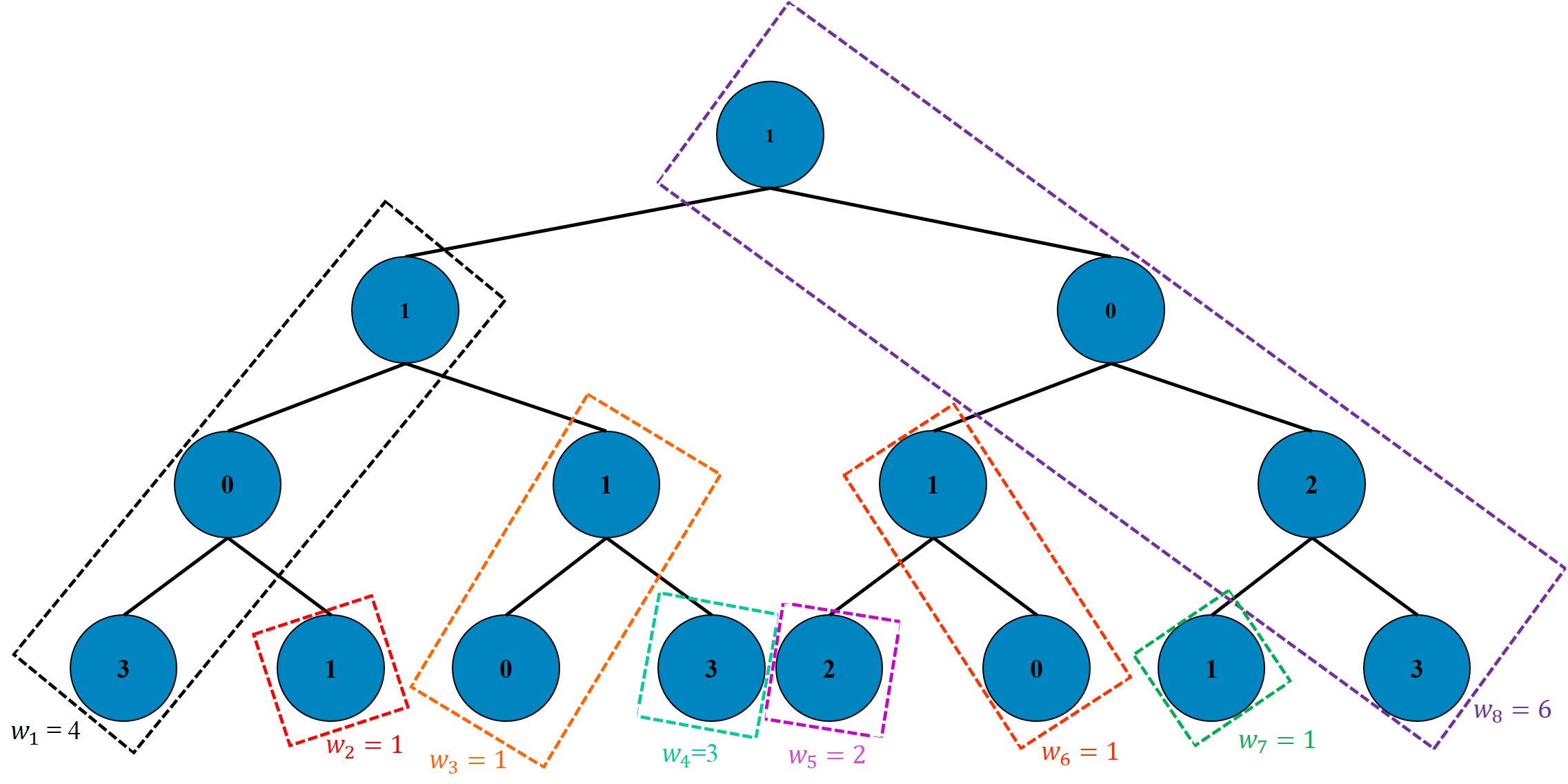}}
\caption{Diagonal cover and Treeplication multiset from Fig.~\ref{fig:8diagonalcover_before} after loss of $l=10$ symbols, no weight zero diagonals.}
\label{fig:8diagonalcover_l10_b1}
\end{figure}

\begin{figure}[htbp]
\centerline{\includegraphics[width=260pt]{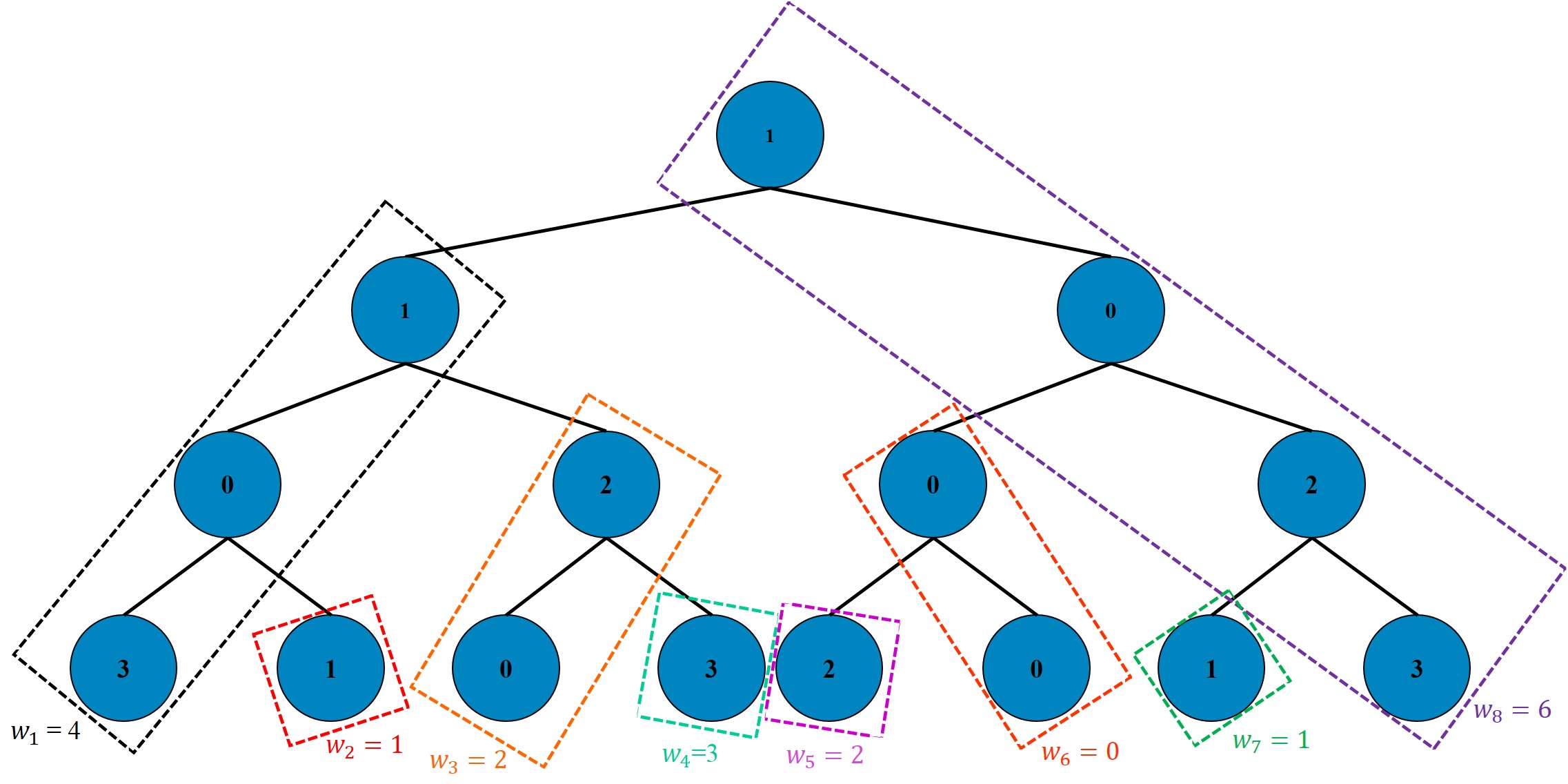}}
\caption{Diagonal cover and Treeplication multiset from Fig.~\ref{fig:8diagonalcover_before} after loss of $l=10$ symbols, one weight zero diagonal.}
\label{fig:8diagonalcover_l10_b0}
\end{figure}

Recall from Definition~\ref{defn:lhealth} that the principal $l$-health of the multiset is defined with respect to loss of $l$ uniformly chosen elements. In the sequel, we assume that the choice of these elements is done by drawing a sequence of $l$ multiset elements (without replacement). Figs.~\ref{fig:8diagonalcover_l10_b1} and \ref{fig:8diagonalcover_l10_b0} illustrate two possible outcomes for the diagonal cover in Fig.~\ref{fig:8diagonalcover_before} after the loss of $l=10$ elements from the sample multiset. In Fig.~\ref{fig:8diagonalcover_l10_b1} no diagonal has weight zero, while in Fig.~\ref{fig:8diagonalcover_l10_b0} $w_6$ dropped to zero.

Toward finding $l$-principal diagonal covers, we next calculate the average probability, under the uniform drawing model, that a diagonal in the cover remains with non-zero weight.
\begin{proposition}
For a Treeplication multiset with weight $\mwgt$ and a diagonal cover of weight profile $W = \{w_j | j \in \{1,\ldots,\kfrag\}\}$, the average probability, over the diagonals of the cover, that the diagonal will have non-zero weight after the loss of $l$ multiset elements, equals

\begin{equation} \label{eq:avrg_prob_cover}
1-{\kfrag}^{-1}\sum_{j = 1}^{\kfrag}\frac{\binom{\mwgt- w_j}{l-w_j} l!}{\mwgt!/(\mwgt-l)!}.
\end{equation}
\end{proposition}

\begin{IEEEproof}
The numerator is the number of drawing sequences that leave the $j$-th diagonal with zero weight. The denominator is the total number of possible drawing sequences. Averaging over the $\kfrag$ diagonals and taking the complement gives~\eqref{eq:avrg_prob_cover}. 
\end{IEEEproof}


\begin{figure}[htbp]
\centerline{\includegraphics[width=150pt]{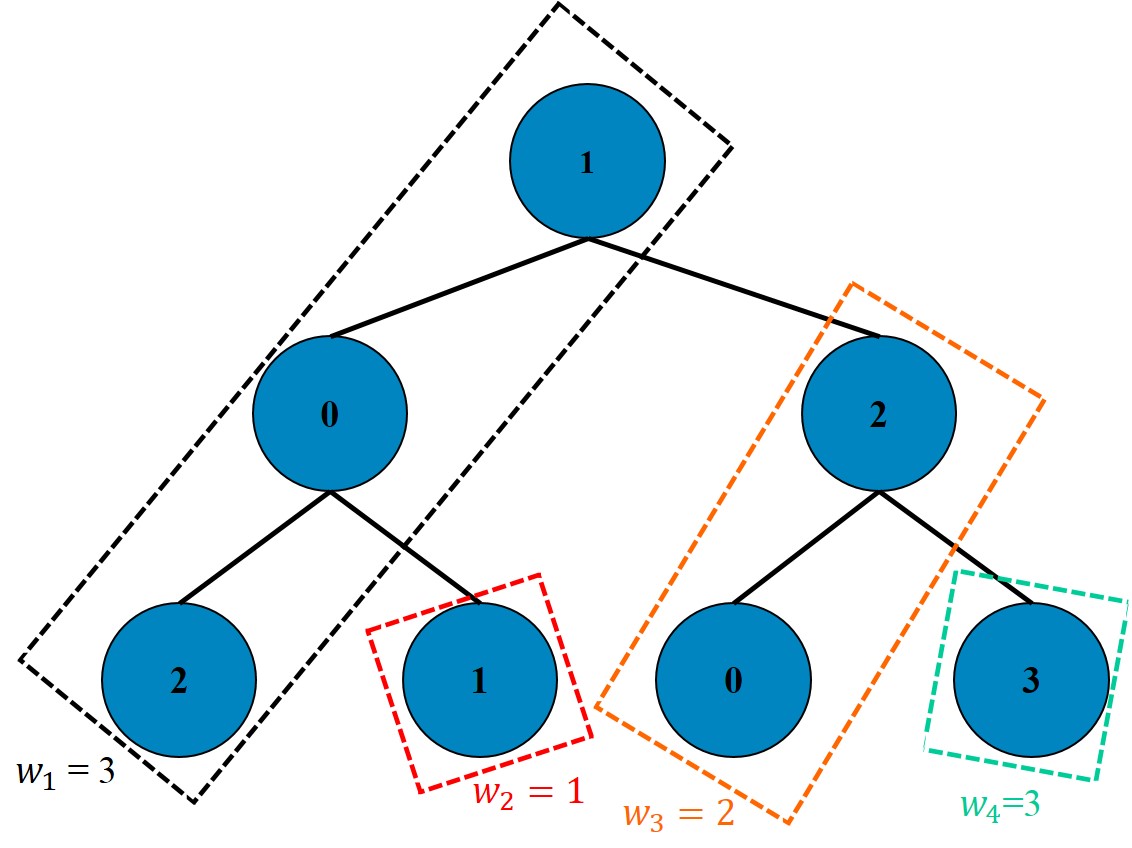}}
\caption{One diagonal cover for a sample Treeplication multiset.}
\label{fig:4diagonalcover_gen}
\end{figure}

\begin{figure}[htbp]
\centerline{\includegraphics[width=150pt]{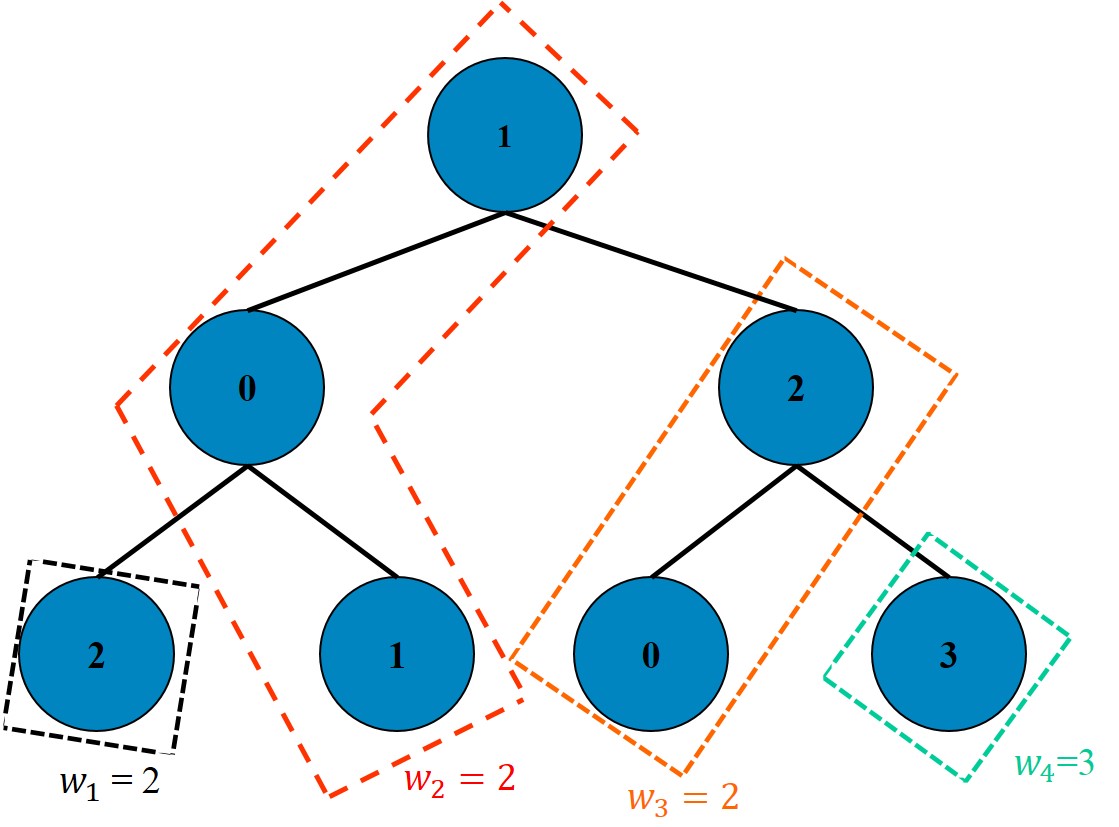}}
\caption{Another diagonal cover for the Treeplication multiset from Fig.~\ref{fig:4diagonalcover_gen} that has a higher average probability in~\eqref{eq:avrg_prob_cover}.}
\label{fig:4diagonalcover_principal}
\end{figure}

Figs.~\ref{fig:4diagonalcover_gen},\ref{fig:4diagonalcover_principal} show two different diagonal covers with respect to the same Treeplication multiset. For $l=3$, the probability~\eqref{eq:avrg_prob_cover} equals $0.890$ for Fig.~\ref{fig:4diagonalcover_gen} and $0.935$ for Fig.~\ref{fig:4diagonalcover_principal}. We later see that the cover in Fig.~\ref{fig:4diagonalcover_principal} is a $l$-principal diagonal for $l=3$ (as well as for all other $l$ values), which means that $0.935$ is maximal for $l=3$ among all possible covers.     

We propose Algorithm~\ref{alg:prindiag} for constructing a diagonal cover for a Treeplication multiset. Algorithm~\ref{alg:prindiag} can be explained in words as growing the diagonals upward, where lower-weight diagonals are preferred when choosing to which diagonal to add a vertex.  

\begin{algorithm}
\caption{Obtain Principal Diagonal Cover}
\begin{algorithmic}[1]

\For {$j \in [1, \kfrag]$}
\State $g_j:= \{(1,j)\}$ // init each diagonal to contain a unique leaf
\EndFor

\For {$i \in [2, d]$}
	\For {$\ell \in [1, 2^{d-i}]$}
		 \State {$y=j:(i-1,2\ell-1) \in g_j $ } // diag. of left child 
		 \State {$z=j:(i-1,2\ell) \in g_j $ } // diag. of right child
			\If{$w_y<w_z$} // add vertex to lighter diagonal
				\State{$g_y = g_y \cup \{(i,\ell)\}$} 
				\Else{}
				\State{$g_z = g_z \cup \{(i,\ell)\}$}
			\EndIf
			
		\EndFor
\EndFor

\State $G:=\{g_j \mid j \in [1,\kfrag]\}$

\State \Return $G$
\end{algorithmic}
\label{alg:prindiag}
\end{algorithm}

\begin{figure}[htbp]
\centerline{\includegraphics[width=150pt]{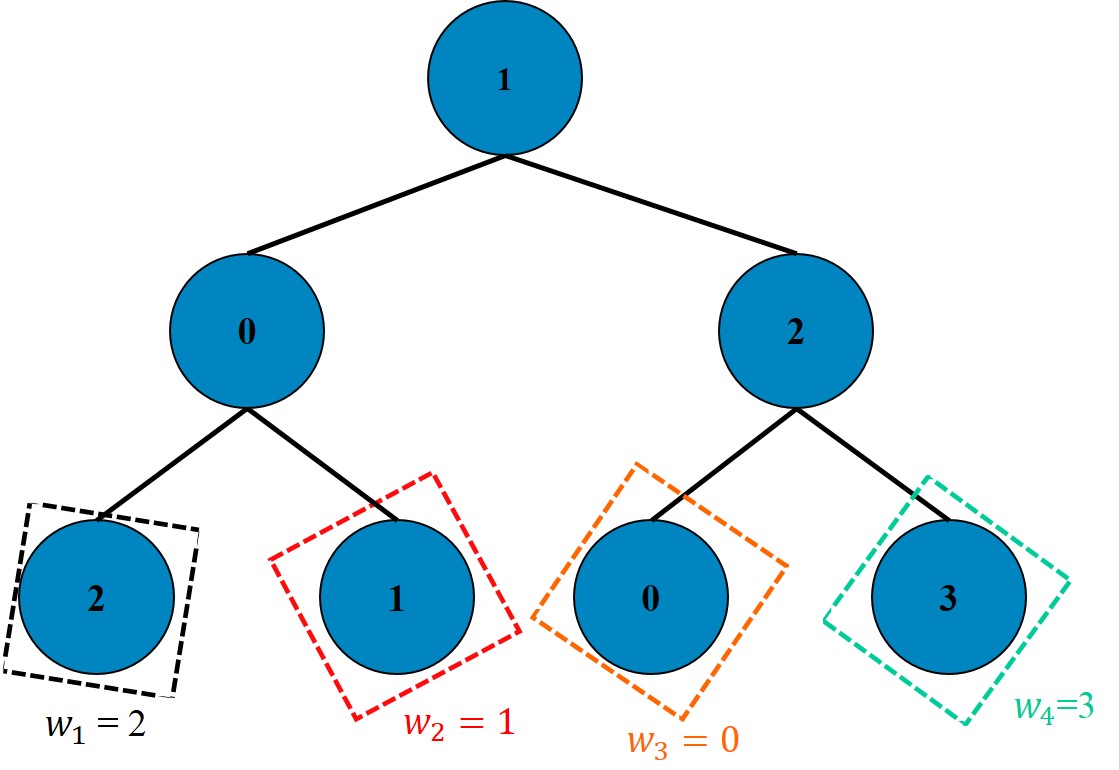}}
\caption{Sample run of Algorithm~\ref{alg:prindiag} after initialization (lines 1-3).}
\label{fig:3tree_step1}
\end{figure}

\begin{figure}[htbp]
\centerline{\includegraphics[width=150pt]{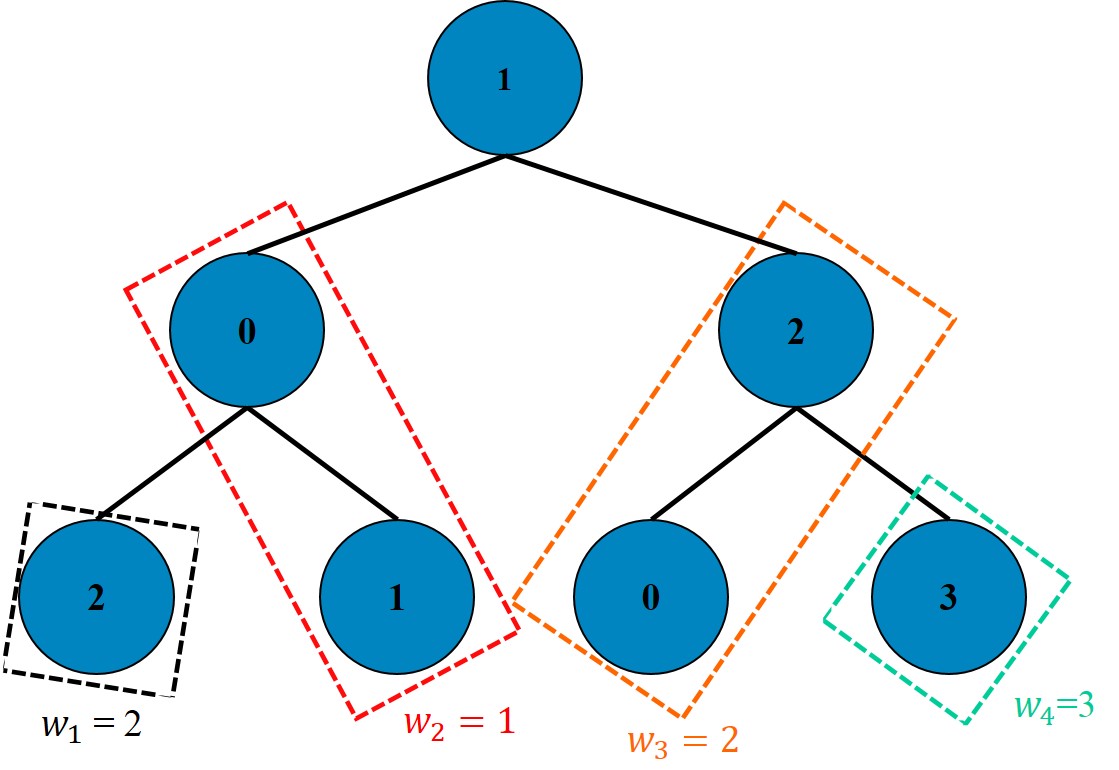}}
\caption{Sample run of Algorithm~\ref{alg:prindiag} after iteration $i=2$. The diagonals chosen to extend upwards are the lower-weight ones between the diagonals of two sibling vertices.}
\label{fig:3tree_step2}
\end{figure}

\begin{figure}[htbp]
\centerline{\includegraphics[width=150pt]{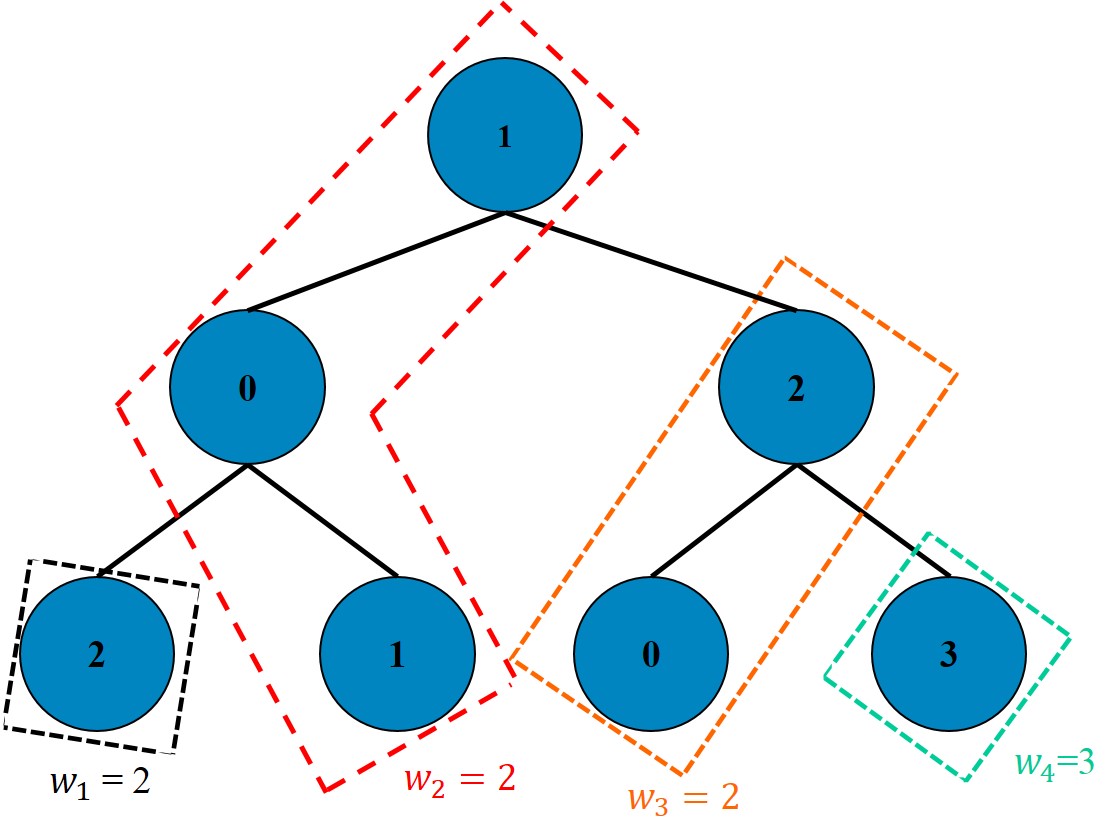}}
\caption{Sample run of Algorithm~\ref{alg:prindiag} after iteration $i=3$.}
\label{fig:3tree_step3}
\end{figure}

Algorithm~\ref{alg:prindiag} assigns each leaf to a different diagonal in line 2 (diagonal initialization), as illustrated in Fig.~\ref{fig:3tree_step1} for the Treeplication multiset in Figs.~\ref{fig:4diagonalcover_gen},\ref{fig:4diagonalcover_principal}. Lines 4 and 5 loop through all non-leaf vertices, where in each iteration a vertex is added to one diagonal (line 9 or 11). The intermediate contents of the diagonals (tracked by the sets $g_1,\ldots,g_k$) are given in Fig.~\ref{fig:3tree_step2} at iteration $i=2$ and in Fig.~\ref{fig:3tree_step3} at iteration $i=3$. Since every tree vertex gets assigned to one diagonal, the output of the algorithm is a diagonal cover. Next we show that the output diagonal cover is in fact a principal diagonal cover. 

\begin{theorem}\label{th:cover_alg}
For any Treeplication-multiset input, Algorithm~\ref{alg:prindiag} returns a diagonal cover that is an $l$-principal diagonal cover for every $l$. 
\end{theorem}

\begin{IEEEproof}
To prove that the algorithm outputs a principal diagonal cover, we need to show that the probability in~\eqref{eq:avrg_prob_cover} is maximized among all possible covers. Since $\kfrag$, $l$ and $\mwgt$ are constants that do not depend on the cover, maximizing the average probability of~\eqref{eq:avrg_prob_cover} is equivalent to minimizing 
\begin{equation} \label{eq:sum_ways_diagonal_loss}
\sum_{j = 1}^{\kfrag}{\binom{\mwgt- w_j}{l-w_j}}.
\end{equation}

Given a diagonal cover, for each non-leaf vertex we distinguish between its child that is assigned to the same diagonal (mate child), and its child that is in a different diagonal (non-mate child). In any cover every non-leaf vertex has one mate and one non-mate child. Algorithm~\ref{alg:prindiag} guarantees the property that for any non-leaf vertex, the sum of vertex weights below it in its diagonal is less than or equal to the weight of the diagonal of its non-mate child. Now we assume by contradiction that a different algorithm outputs a cover with lower sum~\eqref{eq:sum_ways_diagonal_loss} (and, hence, higher probability~\eqref{eq:avrg_prob_cover}). In that output we have at least one vertex in which this property is not met. For this vertex, denote by $q_1$ the sum of vertex weights below it in its diagonal, and by $q_2$ the weight of the diagonal of its non-mate child. By the contradiction assumption we have $q_1>q_2$. We now show that moving this vertex (and all vertices above it in its diagonal) to the diagonal of its non-mate child will result in a lower sum~\eqref{eq:sum_ways_diagonal_loss}. Denote by $x$ the total weight moved in that operation. To prove that, we need to show that
\begin{equation}\label{eq:ineq_binomial_sum_scenarios}
{\binom{\mwgt-q_2}{l-q_2}} + {\binom{\mwgt-q_1-x}{l-q_1-x}} \geq {\binom{\mwgt-q_1}{l-q_1}} + {\binom{\mwgt-q_2-x}{l-q_2-x}},
\end{equation}      
where the inequality is between terms of~\eqref{eq:sum_ways_diagonal_loss} in which the two outputs differ (all other terms are equal and cancel out). To see that ~\eqref{eq:ineq_binomial_sum_scenarios} is true we use elementary combinatorial identities on Pascal's triangle as follows.
Define $\pP_{r}(j)\triangleq {\binom{r+j-1}{r}}$, where $\pP_{r}(j)$ is called the $j$-th element in the $r$-th diagonal of Pascal's triangle. A well-known identity for Pascal's triangle is $\pP_{r}(j) = \sum_{y=0}^{j} \pP_{r-1}(y)$. With that notation, we can write 
\begin{equation}\label{eq:binoms_to_figurate1}
{\binom{\mwgt-q_2}{l-q_2}}-{\binom{\mwgt-q_1}{l-q_1}} = \sum_{y=l-q_1+2}^{l-q_2+1} \pP_{\mwgt-l-1}(y),
\end{equation}
\begin{equation}\label{eq:binoms_to_figurate2}
{\binom{\mwgt-q_2-x}{l-q_2-x}}-{\binom{\mwgt-q_1-x}{l-q_1-x}} = \sum_{y=l-q_1-x+2}^{l-q_2-x+1} \pP_{\mwgt-l-1}(y).
\end{equation}
Another well known property of Pascal's triangle is that elements of its diagonals increase with the argument $j$. This implies 
\begin{equation}\label{eq:ineq_figurate_nums}
\sum_{j=l-q_1+2}^{l-q_2+1} \pP_{\mwgt-l-1}(j) \geq \sum_{j=l-q_1-x+2}^{l-q_2-x+1} \pP_{\mwgt-l-1}(j),
\end{equation}
because the two sums have the same number of summands and the left-hand side is shifted to larger arguments.
From~\eqref{eq:ineq_figurate_nums} and~\eqref{eq:binoms_to_figurate1},\eqref{eq:binoms_to_figurate2} we conclude~\eqref{eq:ineq_binomial_sum_scenarios}, which contradicts the existence of a higher-probability diagonal cover.
\end{IEEEproof}
Since the diagonal cover in Fig.~\ref{fig:4diagonalcover_principal} is the output of Algorithm~\ref{alg:prindiag}, the principal $3$-health of the multiset is now proved to be $0.935$.  

\subsection{Empirical tree-health distribution} 
\begin{figure}[htbp]
\centerline{\includegraphics[width=250pt]{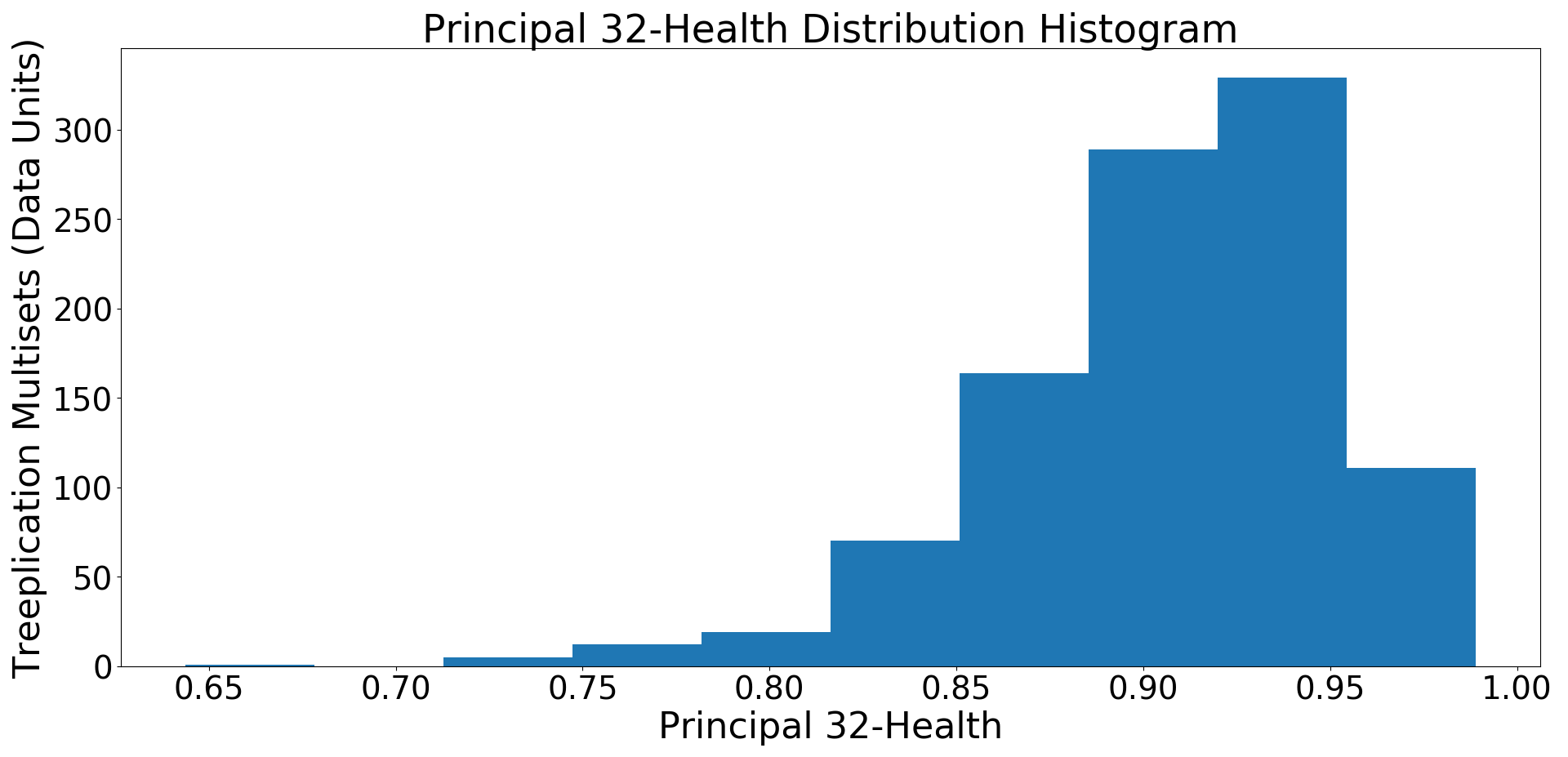}}
\caption{Distribution of the principal $32$-health across $1000$ Treeplication multisets. $\kfrag = 32, \nvert=3\kfrag$.}
\label{fig:healthdist}
\end{figure}

To illustrate the tree-health measure proposed in this section, we show the distribution of principal $l$-healths of $1000$ multisets drawn from the optimal non-uniform selection distribution of Section~\ref{sec:nonuniform}. We take $\kfrag=32$ and multiset size $\nvert=3\kfrag=96$, and plot the distribution of the principal $l$-health, for $l=32$ (after running Algorithm~\ref{alg:prindiag} on each multiset). The results are plotted in the histogram of Fig.~\ref{fig:healthdist}. It can be seen that multisets from the same selection distribution have significant health variability. In practice, if each multiset represents a different data unit stored in the system, we will seek to invest storage resources in increasing the multiset sizes of the data units with the lowest principal $l$-health. In the next section we discuss methods to improve the multiset health in a decentralized way.
 
\section{Distributed Code Augmentation}\label{sec:augment}
When a data unit in the system has low tree health as defined in the previous section, a natural corrective measure is to augment it by adding more code fragments. In this section, we address the problem of deciding {\em which} code fragments to add for this data unit. To fit within decentralized distributed systems, we seek solutions that make those decisions without knowing the full current state of the Treeplication multiset of the data unit. Suppose $\nodeset$ is the set of nodes storing code fragments of a particular data unit. In the sequel, augmenting is done by a node not previously in $\nodeset$ that generates and stores a code fragment while only having access to a subset $\nodesset\subset \nodeset$ of nodes, which we call the {\em accessible nodes}.     
\begin{defn}
Given a subset of accessible nodes of a data unit, we define \textbf{distributed augmentation} as the operation of adding a code fragment using information from accessible nodes.   
\end{defn}
Note that the distributed augmentation operation is divided into first deciding which code fragment to add, and then communicating code fragments from accessible nodes to generate this code fragment. Restricting the set of accessible nodes to be small improves the efficiency in a distributed system, because fewer accessible nodes mean fewer resources used by the node performing augmentation. 

A simple example of distributed augmentation for uncoded replication is augmenting by adding a data fragment chosen from the data fragments in the accessible nodes. For Treeplication, we next propose a scheme that uses the tree structure of the code to augment data units using a small accessible node set $\nodesset$.  

\subsection{Augmenting Treeplication with small accessible node sets}
We now specify a procedure for distributed augmentation where the accessible nodes are those that hold fragments of two sibling vertices and their parent.
\begin{augment}\label{aug:1}
\begin{enumerate}
\item Pick a node $z$ in $\nodeset$
\item Identify the vertex $\vchosen$ stored in $z$ 
\item Take $\nodesset$ to be all nodes storing either $\vchosen$, its parent $\vparent$, or its sibling $\vsibling$.
\item If only $\vchosen$ is present in $\nodesset$, augment by replicating it. Else:
\item Choose the vertex from $\vchosen$,$\vsibling$ with lower weight and augment by replicating it (if exists), or generating it using $\vparent$ (if not).

\end{enumerate}
\end{augment}
Remarks: 1) In step 5 we always choose to augment one of the two siblings $\vchosen$,$\vsibling$, but need the parent $\vparent$ in cases where the sibling $\vsibling$ has zero weight. 2) In case of equal-weight vertices in step 5, we break ties arbitrarily. 

In a real system, a reasonable way to choose $z$ from $\nodeset$ in step 1 of Treeplication Augmentation~\ref{aug:1} is uniformly. The size of the accessible node set $\nodesset$ equals the number of appearances of $\vchosen$,$\vparent$,$\vsibling$ in the multiset, which depends on the chosen $z$. The intuition to augment the lower-weight (weaker) sibling is that it is better than the stronger sibling in improving the probability that the subtree remains decodable locally, and better than the parent in improving the survival probability when the fragment $\vparent$ can be recovered from elsewhere in the full tree.  

Later in the section we compare Treeplication Augmentation~\ref{aug:1} to the following simpler alternative.
\begin{augmentn}\label{augn:1}
\begin{enumerate}
\item Pick a node $z$ in $\nodeset$
\item Identify the vertex $\vchosen$ stored in $z$ 
\item Replicate $\vchosen$ 
\end{enumerate}
\end{augmentn}

In Augmentation by replication we simply copy the code fragment from the node we picked in line 1 to the new node entering $\nodeset$. This procedure can apply to both Treeplication and standard replication, where in the latter $\vchosen$ is always a data fragment.   
 
\begin{figure}[htbp]
\centerline{\begin{subfigure}{.2\textwidth}
  \centering
  \includegraphics[width=.6\linewidth]{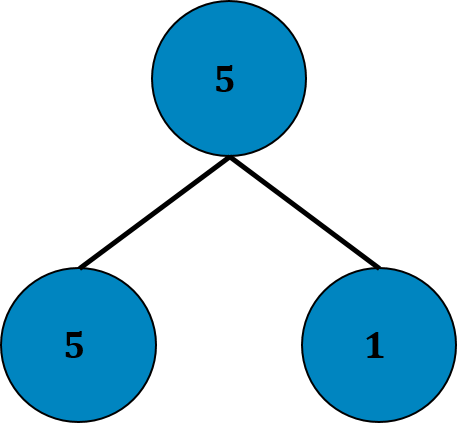}
  \caption{Before augmentation.}
\end{subfigure}
\begin{subfigure}{.2\textwidth}
  \centering
  \includegraphics[width=.6\linewidth]{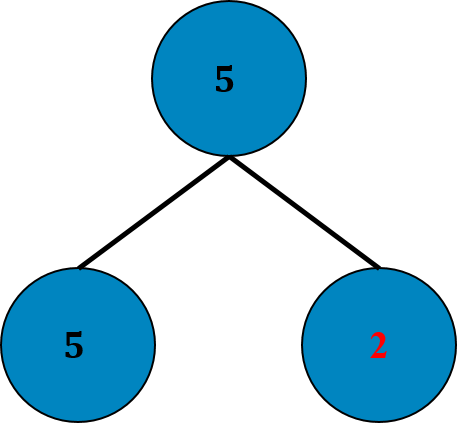}
  \caption{After augmentation.}
\end{subfigure}}
\caption{Treeplication Augmentation~\ref{aug:1} for a $k=2$ tree and a sample Treeplication multiset.}
\label{fig:aug_lightest_example}
\end{figure}

An example of Treeplication Augmentation~\ref{aug:1} is presented in Fig.~\ref{fig:aug_lightest_example} for a $k=2$ tree and a sample Treeplication multiset. Because in Fig.~\ref{fig:aug_lightest_example}a the right leaf has lower weight than the left leaf, the former is chosen to be augmented (Fig.~\ref{fig:aug_lightest_example}b). Before augmentation, the probability that the multiset remains decodable (survives) after loss of $9$ nodes is $0.64$, and increasing to $0.91$ after augmenting by one code fragment. If we chose the other (stronger) leaf for augmentation, the survival probability would only go up to $0.86$.

We propose Treeplication Augmentation~\ref{aug:1} mainly as an example how with a small accessible set $\nodesset$ we can improve survivability over Augmentation by replication. There are many possible generalizations and enhancements of Treeplication Augmentation~\ref{aug:1} that can further improve performance. For example, it is possible to consider bigger subtrees in the augmentation choice, and solve interesting optimization problems to maximize global survival probability given local subtree information.    

\subsection{Empirical study: augmentation in node birth-death processes} 
To study and compare augmentation schemes in distributed systems, we next define a dynamic system setup where augmentation affects the long term survival of data units. To that end, we use a discrete-time {\em birth-death process} to model the dynamics of the data-unit's node set. At each time instant, an event of either birth (addition of a node) or death (removal of a node) occurs in the data unit. If birth is drawn, adding a node invokes an augmentation operation. If death is drawn, a randomly selected node is chosen to be removed from $\nodeset$ of that data unit. We restrict ourselves to balanced processes, where birth and death occur each with probability $0.5$. In general, birth-death processes may have time instants where neither birth nor death occurs, but for the purpose of comparing augmentation schemes these time instants are not interesting. We use the term {\em generation} to define the state of the data unit following the node removal in death instants and augmentation in birth instants. We define the {\em data loss} event for the data unit as the first generation where the data unit becomes non-decodable. Since data-loss is irreversible, the process terminates immediately after. In our results the number of generations a data unit {\em survives} is the number of process instants before the data-loss event. 

\begin{figure}[htbp]
\centerline{\includegraphics[width=400pt]{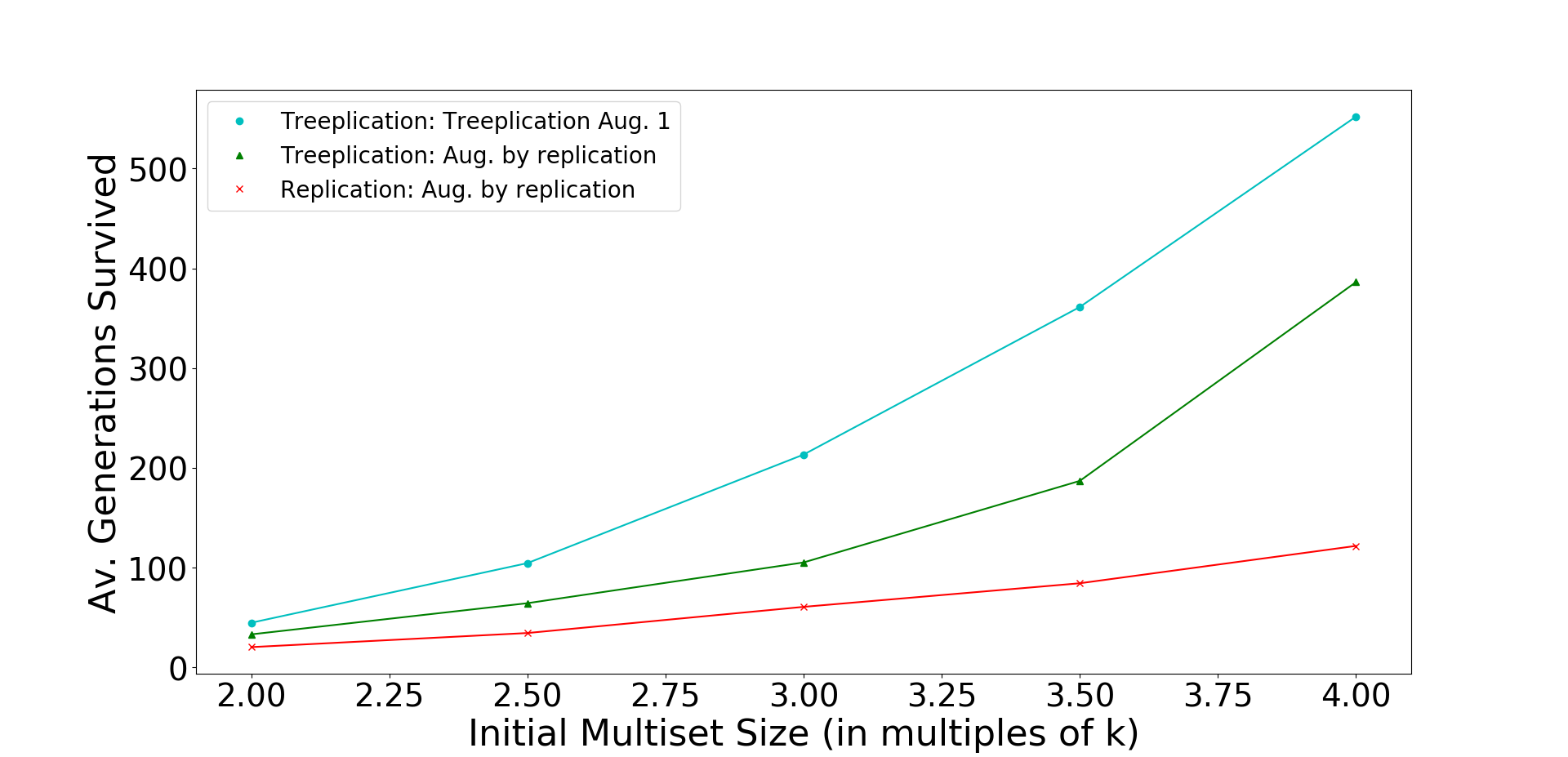}}
\caption{Augmentation in a birth-death process. From bottom to top: replication with Augmentation by replication, Treeplication with Augmentation by replication, and Treeplication with Treeplication Augmentation~\ref{aug:1}. $k = 32$.}
\label{fig:gens_tree_rep}
\end{figure}

In Fig.~\ref{fig:gens_tree_rep} we compare the average number of generations survived in three different setups: 1) replication with Augmentation by replication, 2) Treeplication with Augmentation by replication, and 3) Treeplication with Treeplication Augmentation~\ref{aug:1}. For each setup we simulated the birth-death process, and plotted the average number of generations survived across 3000 runs. In each run we randomly drew the $0$-th generation of the data unit: for Treeplication using the optimal non-uniform distribution (from Section~\ref{sec:nonuniform}), and for replication uniformly from the data fragments. Fig.~\ref{fig:gens_tree_rep} demonstrates that Treeplication fares better than replication also in the dynamic regime, and more interestingly, that Treeplication Augmentation~\ref{aug:1} improves significantly over Augmentation by replication. 

\section{Conclusion}
We have shown that Treeplication codes combine the strength of erasure codes in recoverability with replication-like access performance. The tree structure of Treeplication allows deriving exact recursive expressions toward the analysis and optimization of the code under random-multiset models.  It is an interesting open problem to generalize the code structure beyond a binary tree while still maintaining the algorithmic and analysis capabilities we demonstrated for Treeplication.  


\end{document}